\documentclass{mn2e}

\usepackage{amsfonts}
\usepackage{amsmath}
\usepackage{graphicx}

\title[Curvature in the color-magnitude relation but not in color$-\sigma$]
      {Curvature in the color-magnitude relation but not in color$-\sigma$: 
       Major dry mergers at $M_*> 2\times 10^{11}M_\odot$?}

\author[M. Bernardi et. al.]
{Mariangela Bernardi$^1$\thanks{E-mail: bernardm@physics.upenn.edu}, 
 Nathan Roche$^2$, Francesco Shankar$^{3}$ \& Ravi K. Sheth$^{1,4}$\\
 $^1$Department of Physics \& Astronomy, University of Pennsylvania, 
      209 S. 33rd St., Philadelphia, PA 19104, USA\\
 $^2$ Dipartimento di Astronomia, Universit\'{a} degli Studi di Bologna, 
      via Ranzani 1, I-40127 Bologna, Italy\\
 $^3$ Max-Planck-Instit\"{u}t f\"{u}r Astrophysik,
      Karl-Schwarzschild-Str. 1, D-85748, Garching, Germany \\
 $^4$ Center for Particle Cosmology, University of Pennsylvania, 
      209 S. 33rd St., Philadelphia, PA 19104, USA}
\begin{document}
\pagerange{\pageref{firstpage}--\pageref{lastpage}}

\maketitle 

\label{firstpage}

\begin{abstract}
The color-magnitude relation of early-type galaxies differs slightly 
but significantly from a pure power-law, curving downwards at low 
and upwards at large luminosities ($M_r>-20.5$ and $M_r<-22.5$, 
respectively).  This remains true of the color-size relation, and is 
even more apparent with stellar mass ($M_* < 3\times10^{10} M_\odot$ 
and $M_* > 2\times 10^{11}M_\odot$, respectively).  
The upwards curvature at the massive end does not appear to be due to stellar 
population effects.  
In contrast, the color-$\sigma$ relation
is well-described by a single power law.  
Since major dry mergers change neither the colors nor $\sigma$, but they 
do change masses and sizes, 
the clear features observed in the scaling relations with $M_*$, 
but not with $\sigma > 150 \, {\rm km~s}^{-1}$,  suggest that 
$M_* > 2\times 10^{11}M_\odot$ is the scale above which major dry
mergers dominate the assembly history.  


We discuss three models of the merger histories since $z\sim 1$ which are 
compatible with our measurements.  In all three models, dry mergers are 
responsible for the flattening of the color-$M_*$ relation at 
$M_* > 3\times10^{10} M_\odot$ -- wet mergers only matter at smaller masses.  
At $M_* > 2\times 10^{11}M_\odot$, the merger histories in one model are 
dominated by major rather than minor dry mergers.
In another, although both major and minor mergers occur at the high mass 
end, the minor mergers contribute primarily to the formation of the ICL, 
rather than to the stellar mass growth of the central massive galaxy.  
This model attributes the fact that $\alpha < 1$ in the 
scaling $M_*\propto M_{\rm dyn}^\alpha$, to the formation of the ICL.
A final model assumes that the bluest objects today were assembled by
minor dry mergers of the bluest (early-type) objects at high redshift, 
whereas the reddest objects were assembled by a mix of major 
and minor dry mergers.  In this model, the scatter of the color-magnitude 
relation should increase with redshift, and the dependence on environment 
should also be more pronounced at higher redshift:  more clustered objects 
should be redder.  
Similar measurements of these relations at high redshift will provide 
further valuable constraints on the mass scale at which major dry mergers 
dominate the assembly history. 

\end{abstract}


\section{Introduction}
The colors of early-type galaxies are tightly correlated with their 
luminosities (Sandage \& Visvanathan 1978).  The mean relation is 
well-described by a single-power law whose slope evolves little out 
to $z\sim 1$ (e.g. Kodama et al. 1998; Mei et al. 2009).
This, and the small scatter around the mean relation, are thought 
to imply that the stellar populations in these objects are old 
(e.g. Bower et al. 1992; Bernardi et al. 2003b,c), although the 
total stellar metallicity, the $\alpha$-elements-to-iron abundance 
ratio, and light-weighted age, all increase along the relation 
(e.g. Bernardi et al. 2006; Gallazzi et al. 2006).  
Data sets are now large enough that significant departures from 
simple power laws can be detected:  the mean color-magnitude relation 
appears to be steeper at faint luminosities (e.g. Baldry et al. 2004; 
Graham 2008; Skelton et al. 2009).  
This change in slope is thought to indicate that the mechanism by 
which the stars were assembled into a single object is different at 
low luminosities than at higher ones.  

However, different morphological types define different 
color-magnitude relations.  Since the mix of morphological types is 
a strong function of luminosity, it is possible that the observed 
curvature is really due to morphology, rather than to a change in 
formation histories at fixed morphology.  
Unfortunately, it is difficult to select large samples of a given 
morphological type that are pure.  In what follows, we compare the 
color-magnitude relation obtained from a number of different ways 
of defining an early-type sample.  We argue that while the steepening 
of the relation at faint luminosities may be affected by morphological 
effects, it appears to be present even in relatively pure samples of 
ellipticals -- this may arise from the fact that dwarf and giant 
ellipticals are known to be different in other ways.  
However, we also show that, at the very highest luminosities, 
$M_r < -22.5$, the relation steepens again.  
In a companion paper, Bernardi et al. (2010b) show that this 
steepening occurs on the same scale where the size-luminosity and 
velocity dispersion-luminosity relations steepen and flatten, 
respectively (Bernardi et al. 2007).  In addition, the trend for 
axis-ratio and color-gradient to increase with luminosity, reverse 
on this scale (Bernardi et al. 2008; Roche et al. 2010).  
When expressed in terms of stellar mass, the relevant scale is 
$M_* = 2\times 10^{11}M_\odot$.  

Section~\ref{sample} describes the SDSS sample, and a number of 
ways for selecting early-types from it.  
Section~\ref{bent} presents the associated color-magnitude relations, 
and shows that the trends we see are even more pronounced if we replace 
luminosity with stellar mass.
It also shows that, in contrast, the color-$\sigma$ relation is 
well-described by a single power law over essentially the entire 
range of $\sigma$.
Section~\ref{sec:ageZ} shows that the curvature is not due to 
stellar population effects.   
Section~\ref{sec|theory} compares our empirical results with simple 
models.  While these toy models are not intended to provide a precise 
quantification of the color evolution, they provide a useful framework 
within which to discuss our measurements.   
A final section summarizes our findings, and discusses what 
they suggest about how the formation and assembly of early-type 
galaxies depend on mass and redshift. 

Appendix~\ref{2gs} describes a way of selecting early-types which 
exploits the fact that galaxy properties are approximately bimodal; 
Appendix~\ref{fukugita} contrasts this with selection based on 
eyeball classifications of morphology.  
A number of tests of systematics -- robustness to changes in the 
scale on which rest-frame color is measured or inferred (color 
gradients and $k+e$ corrections) -- are described in 
Appendix~\ref{systematics}.  
A final Appendix provides details of the expected changes to 
galaxy sizes and velocity dispersions if galaxy mergers occur along 
parabolic orbits and conserve mass and energy.  

Where necessary we assume a flat background geometry that is 
dominated at the present time by a cosmological constant 
$\Lambda_0=1-\Omega_0$, where $\Omega_0 = 0.3$ is the background 
density in units of the critical density, with Hubble constant 
$H_0 = 70$~km~s$^{-1}$Mpc$^{-1}$.

\section{Sample}\label{sample}

\subsection{Data}\label{data}

In what follows, we will use the luminosities, colors, velocity 
dispersions and stellar masses of a magnitude limited sample of
$\sim 250,000$ SDSS galaxies with $14.5 < m_{\tt Pet} < 17.5$ 
in the $r-$band, selected from 4681 deg$^2$ of sky.  In this band, 
the absolute magnitude of the Sun is $M_{r,\odot} = 4.67$. 

We use the {\tt cmodel} magnitudes as well as the {\tt Petrosian} 
and {\tt model} $g-r$ colors output by the SDSS database. 
The {\tt cmodel} magnitude is a very crude disk+bulge magnitude 
which has been seeing-corrected.  
Rather than resulting from the best-fitting linear combination of an 
exponential disk and a deVaucouleur bulge, the {\tt cmodel} magnitude 
comes from separately fitting exponential and deVaucouleur profiles 
to the image, and then combining these fits by finding that linear 
combination of them which best-fits the image (see Bernardi et al. 2010a
for more discussion).  The analysis which follows does not depend
on whether one uses {\tt cmodel} or {\tt Petrosian} magnitudes.  
({\tt Petrosian} magnitudes are not seeing corrected, and they 
underestimate the total light in a deVaucouleurs profile by about 
0.05~mags.)

However, choosing {\tt model} rather than {\tt Petrosian} colors 
does matter, because of color-gradients:  the {\tt Petrosian} 
color is associated with a larger scale, and so is typically bluer.
For faint galaxies, the {\tt model} colors have higher signal-to-noise 
ratio than do the {\tt Petrosian} colors.
  
We apply $k$- and {\it evolution}-corrections to the luminosities 
and colors.  We use $k$-corrections from Blanton \& Roweis (2007), 
which are based on fitting templates to the observed colors.  Because 
these are suspect at the bright end (Bernardi et al. 2010a argue 
that they assume younger stellar populations than may be realistic), 
we also explore spectral based $k$-corrections from Roche et al. (2009).  
Our evolution correction depends on the $k$-correction:  
we make high redshift objects fainter by $0.9z$ ($r-$band) and 
redder by $0.15z$ for Roche et al. $k$-corrections, 
and by $1.3z$ but with negligible color evolution correction for 
Blanton \& Roweis. 
See Section~\ref{colgrad} and Figure~\ref{gmrEvolM} for more discussion.

We also use the concentration index $C_r$, which is the ratio of the 
scale which contains 90\% of the Petrosian light in the $r$-band to that 
which contains 50\%.  
Finally, we use the velocity dispersions and stellar masses of these 
objects as described in Bernardi et al. (2010a). The stellar masses 
were computed following Bell et al. (2003), who report that, at
 $z=0$, $\log_{10}(M_*/L_r)_0 = 1.097\,(g - r)_0 - 0.406$, 
where the zero-point depends on the IMF (see their Appendix~2 and Table~7).
We calibrate to a Chabrier IMF.  (See Table~2 in Bernardi et al. 2010a for 
how to transform between different IMFs.  Bernardi et al. also report a 
detailed comparison between the different ways of computing stellar masses 
and their biases -- see their discussion of the stellar mass function and 
their Appendix~A.)  In Section~\ref{sec:ageZ} we make use of age and 
metallicity estimates for the objects in our sample.  These come from 
Gallazzi et al. (2005), and are based on absorption line features in the 
spectra.  

\subsection{Sample selection}
In this paper we are interested in early-type galaxies.
The light profiles of such galaxies are more centrally concentrated, 
so they are expected to have larger values of {\tt $C_r$}.  
Two values are in common use:  a more conservative {\tt $C_r$}$\ge 2.86$ 
(e.g. Nakamura et al. 2003; Shen et al. 2003) and a more cavalier 
{\tt $C_r$}$\ge 2.6$ (e.g. Strateva et al. 2001; Kauffmann et al. 2003; 
Bell et al. 2003; Skelton et al. 2009).
We can also select early-type galaxies following Hyde \& Bernardi (2009),
who use a combination of photometric features (a revised version of 
Bernardi et al. 2003a): 
i.e. {\tt fracDev} $= 1$ in $g$- and $r$-, $r$-band $b/a > 0.6$ and
log$_{10} (r_{e,g}/r_{e,r}) < 0.15$. This last condition is essentially 
a cut on color gradient (Roche et al. 2010). 

Recently, Bernardi et al. (2010a) have shown that requiring concentration 
indices $C_r\ge 2.6$ selects a mix in which E+S0+Sa's account for about 
two-thirds of the objects; if $C_r\ge 2.86$ instead, 
then two-thirds of the sample comes from E+S0s; whereas Es alone account 
for more than two-thirds of a sample selected following Hyde \& Bernardi (2009) 
(see Figures~11 and~12, and Table~3 of Bernardi et al. 2010a). 
E's alone account for 
about 40\%, 50\% and 75\% of the total stellar mass in samples 
selected in these three ways. 
In Appendix~\ref{fukugita}, we also present results from a small subset 
of this dataset for which eye-ball classifications of morphology are 
available (from Fukugita et al. 2007).

There is a third method used to select early-type samples from the SDSS
in addition to direct eye-ball classifications (e.g. Fukugita et al. 2007; 
Lintott et al. 2008) and to the two common automated ways introduced above 
(i.e. concentration index and Hyde--Bernardi). This is based on the 
fact that the color-magnitude relation is bimodal 
(e.g. Baldry et al. 2004; Blanton et al. 2005) at least out to redshifts 
of order unity (Willmer et al. 2006).  This bimodality has sometimes 
been used as a simple way to select red sequence galaxies.  Typically, 
one selects objects which lie redward of a straight color cut, or 
redward of a line which lies below, but parallel to, the red sequence
(e.g. Zehavi et al. 2005; Blanton \& Berlind 2007).  
The resulting sample is then treated as though it is comprised of 
early-types, even though it can contain a substantial fraction of 
edge-on spirals (Mitchell et al. 2005; Bernardi et al. 2010a).  
Although a cut on axis ratio can remove such objects 
(Bernardi et al. 2010a), this simple extra step is almost never taken. 

We select a sample using `bimodality' as follows.  
We first divide the full galaxy sample into narrow bins in luminosity.  
We then model the color distribution in each luminosity bin as the sum 
of two Gaussian components.  The means and rms values of the two 
Gaussians, obtained by fitting the model to the data, give the red and 
blue sequences and their scatter; the amplitudes of the Gaussians give 
the fraction of galaxies in each component (e.g. Baldry et al. 2004; 
Skibba \& Sheth 2009).  Appendix~\ref{2gs} provides details, and 
argues that the double-Gaussian decomposition correctly assigns the 
reddest objects at intermediate and low luminosities to the blue sequence.  
The means and rms of the two Gaussians and the fraction of galaxies in 
each component are listed in Table~\ref{gmrMredblue}.


\begin{figure}
 \centering
 \includegraphics[width=0.95\hsize]{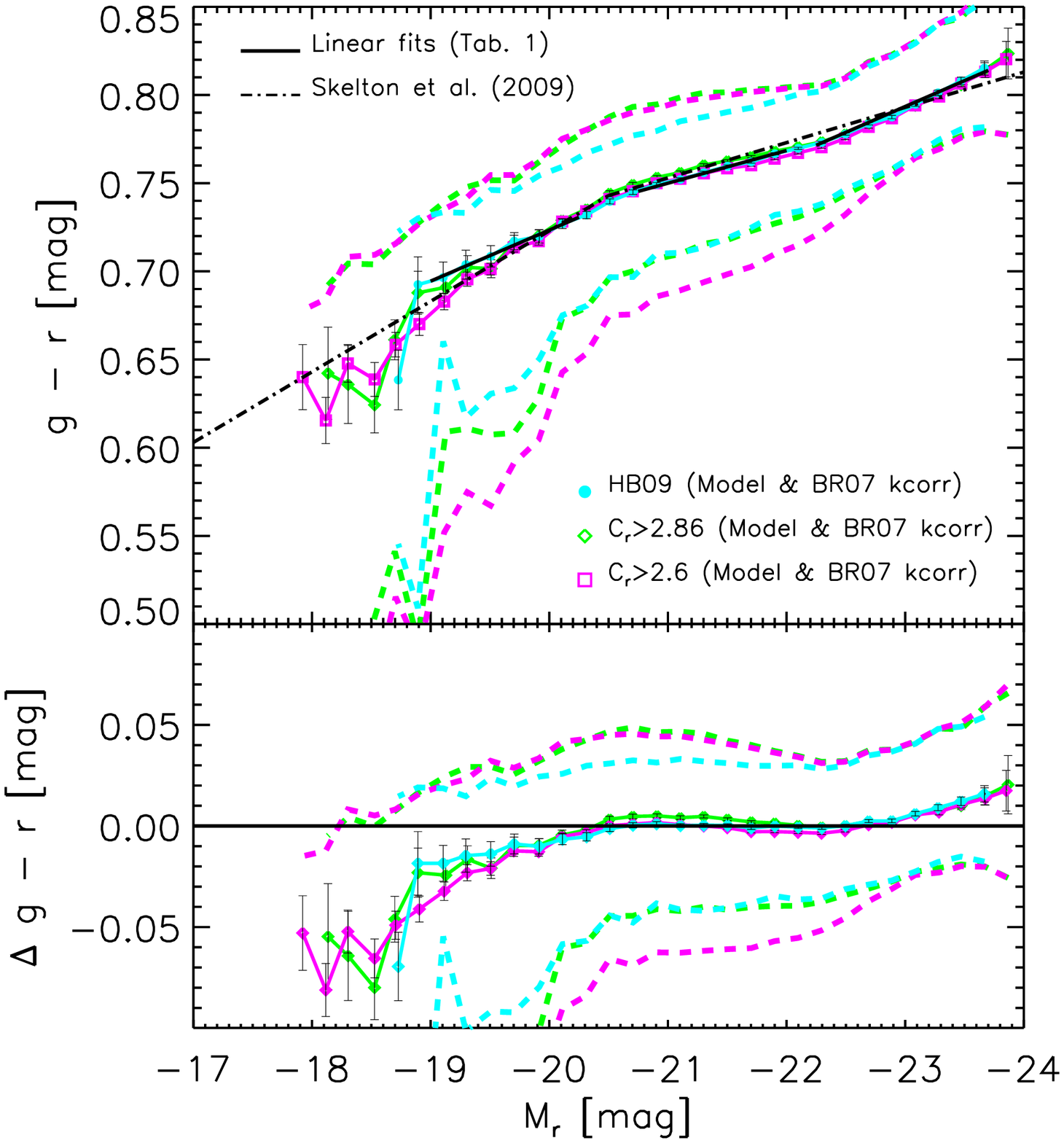}
 \caption{Red sequence defined by various samples (as labeled) when 
          {\tt model} colors and {\tt cmodel} magnitudes are used and 
          the $k$-correction is from Blanton \& Roweis (2007).  
          Top:  Symbols with error bars show the mean $g-r$ for 
          bins in $M_r$, and dashed lines show the rms scatter around 
          this mean, for the different samples.
          Thick solid lines show the three regimes (fits are reported 
          in Table~\ref{gmrMtable}) in the sample which is selected 
          following Hyde \& Bernardi (2009). 
          Dot-dashed lines show the steep and shallow slopes for the 
          faint and bright ends of this relation measured by 
          Skelton et al. (2009) on a sample selected with $C_r > 2.6$ and
          at $z<0.06$.  
          Bottom:  Same as top, except now, to reduce the dynamic
          range, a mean trend has been subtracted from the colors.  
          Plot shows $g-r - (0.361 - 0.019\,M_r)$ versus $M_r$:  the 
          reduction in dynamic range highlights the curvature in the 
          relation.   }
 \label{gmrModel}
\end{figure}

\begin{figure*}
 \centering
 \includegraphics[width=0.475\hsize]{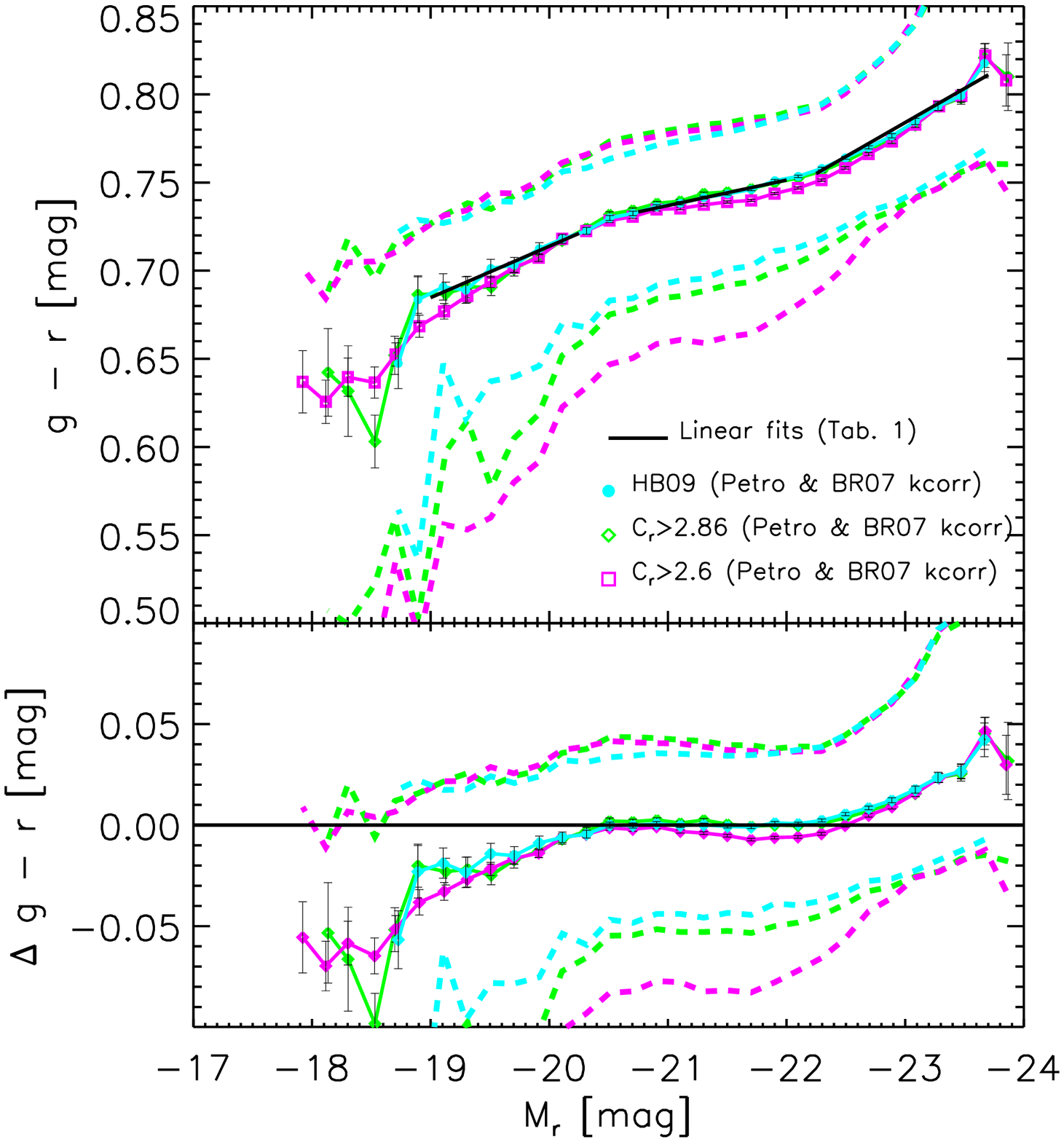}
 \includegraphics[width=0.475\hsize]{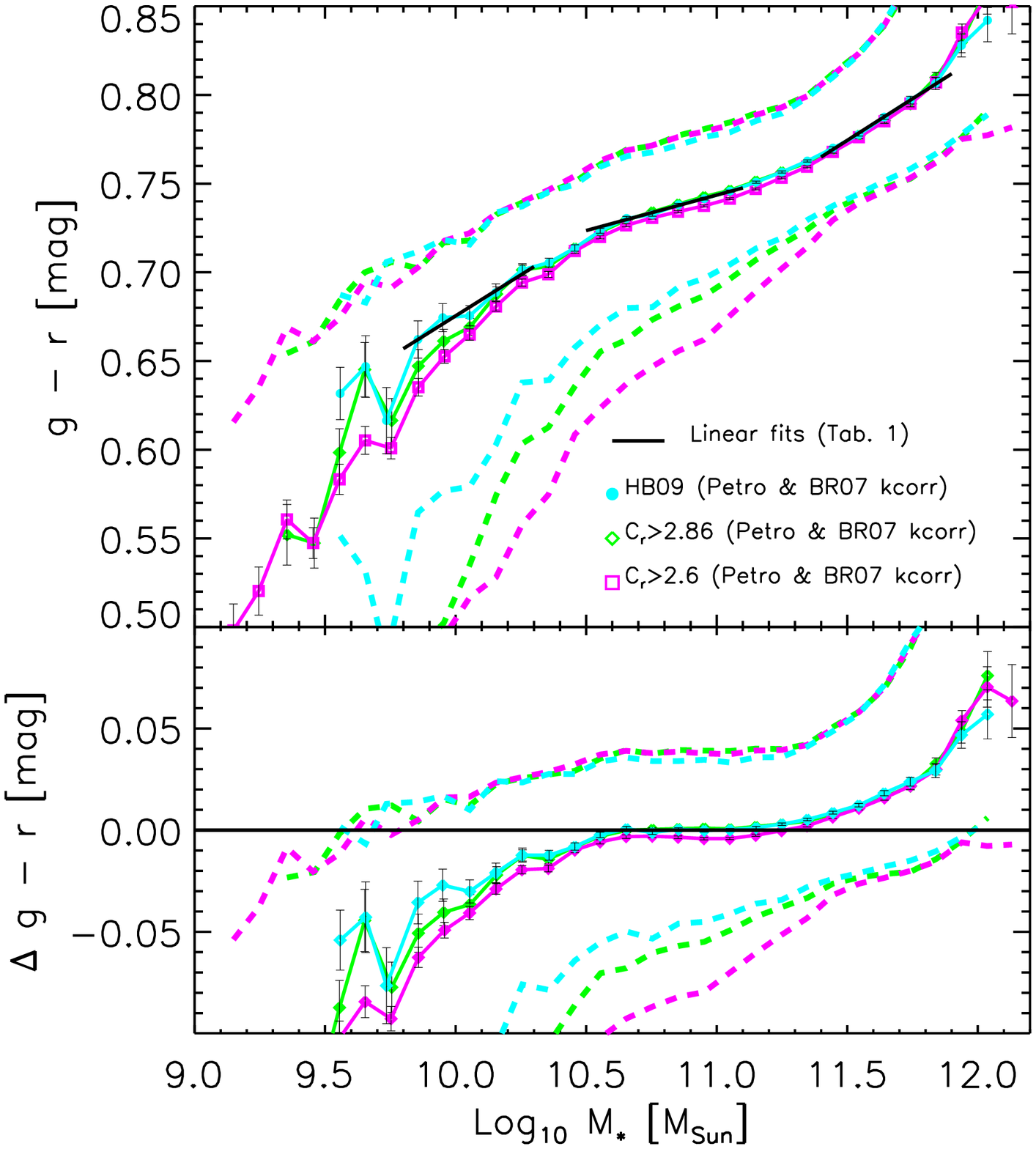}
 \caption{Red sequence defined by various samples (as labeled) when 
          {\tt Petrosian} colors and magnitudes are used and 
          the $k$-correction is from Blanton \& Roweis (2007).  
          Top left:  Symbols with error bars show the mean $g-r$ for 
          bins in $M_r$, and dashed lines show the rms scatter around 
          this mean, for the different samples.
          Thick solid lines show the three regimes (fits are reported 
          in Table~\ref{gmrMtable}) in the sample which is selected 
          following Hyde \& Bernardi (2009). 
          Bottom left:  Same as top panel, except now, to reduce the dynamic
          range, a mean trend has been subtracted from the colors.  
          Plot shows $g-r - (0.434 - 0.014\,M_r)$ versus $M_r$:  the 
          reduction in dynamic range highlights the curvature in the 
          relation.   
          Top right panel shows a similar analysis of $\langle g-r|M_*\rangle$, 
          where colors and stellar masses are derived from {\tt Petrosian} 
          quantities.  The bottom right panel shows 
          $g-r - (0.303 + 0.040 \log_{10} M_*/M_\odot)$ versus $M_*$.}
 \label{gmrEtypes}
\end{figure*}

\section{Curvature}\label{bent}

\subsection{Curvature in the red sequence}
Figure~\ref{gmrModel} shows that the different ways of selecting 
early-type samples mentioned above (cuts in $C_r$, or following 
Hyde \& Bernardi 2009) produce almost indistinguishable 
color-magnitude relations.  
This is remarkable, given that the mean relation they define is not 
a simple power law.  Rather, it bends downward at low luminosities, 
and upward at high luminosities, while being relatively flat at 
intermediate luminosities.  The changes in slope occur around 
$M_r=-20.5$ and $M_r=-22.5$~mags.  
Table~\ref{gmrMtable} quantifies the slopes.  

\begin{figure*}
 \centering
 \includegraphics[width=0.475\hsize]{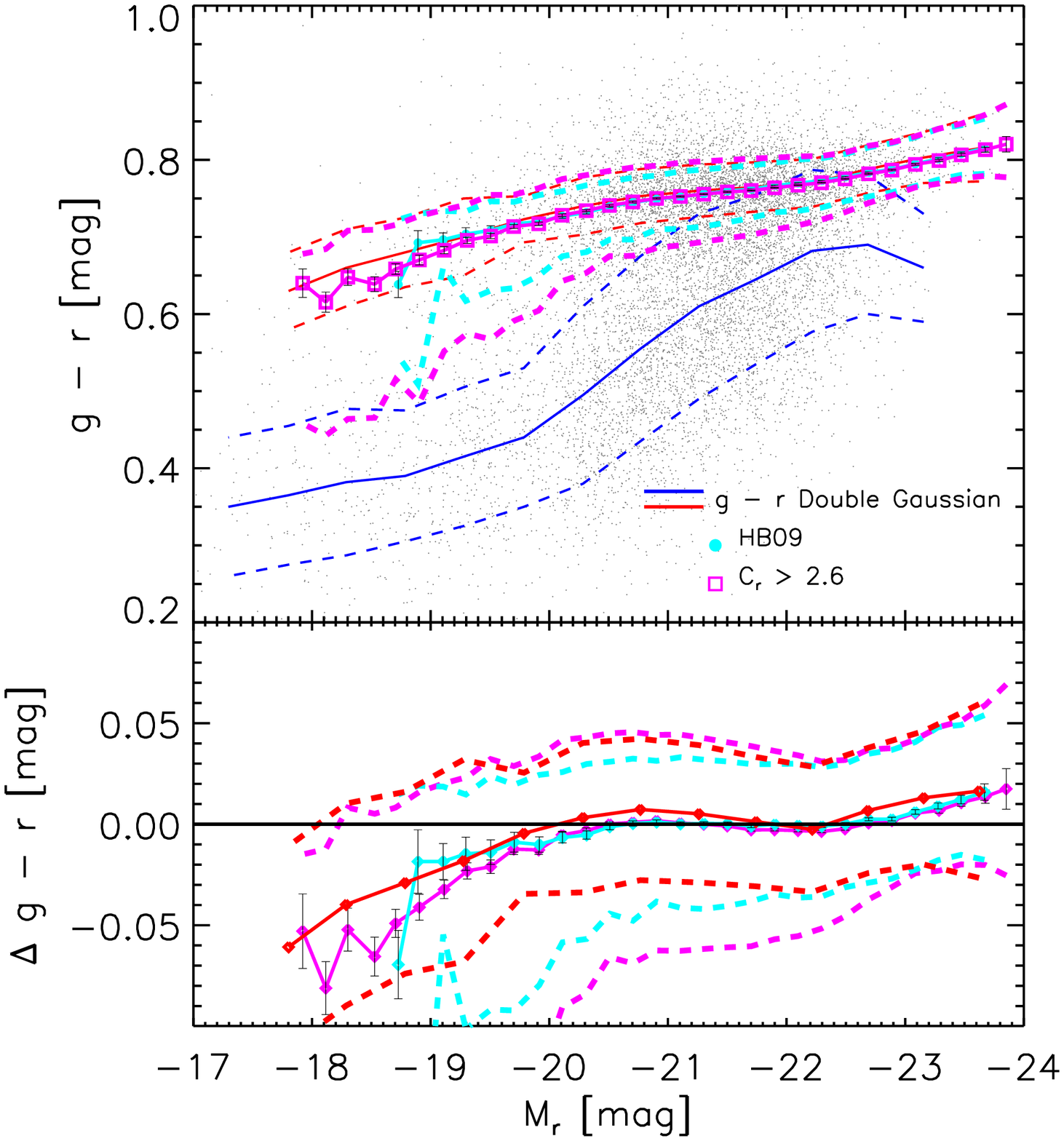}
 \includegraphics[width=0.475\hsize]{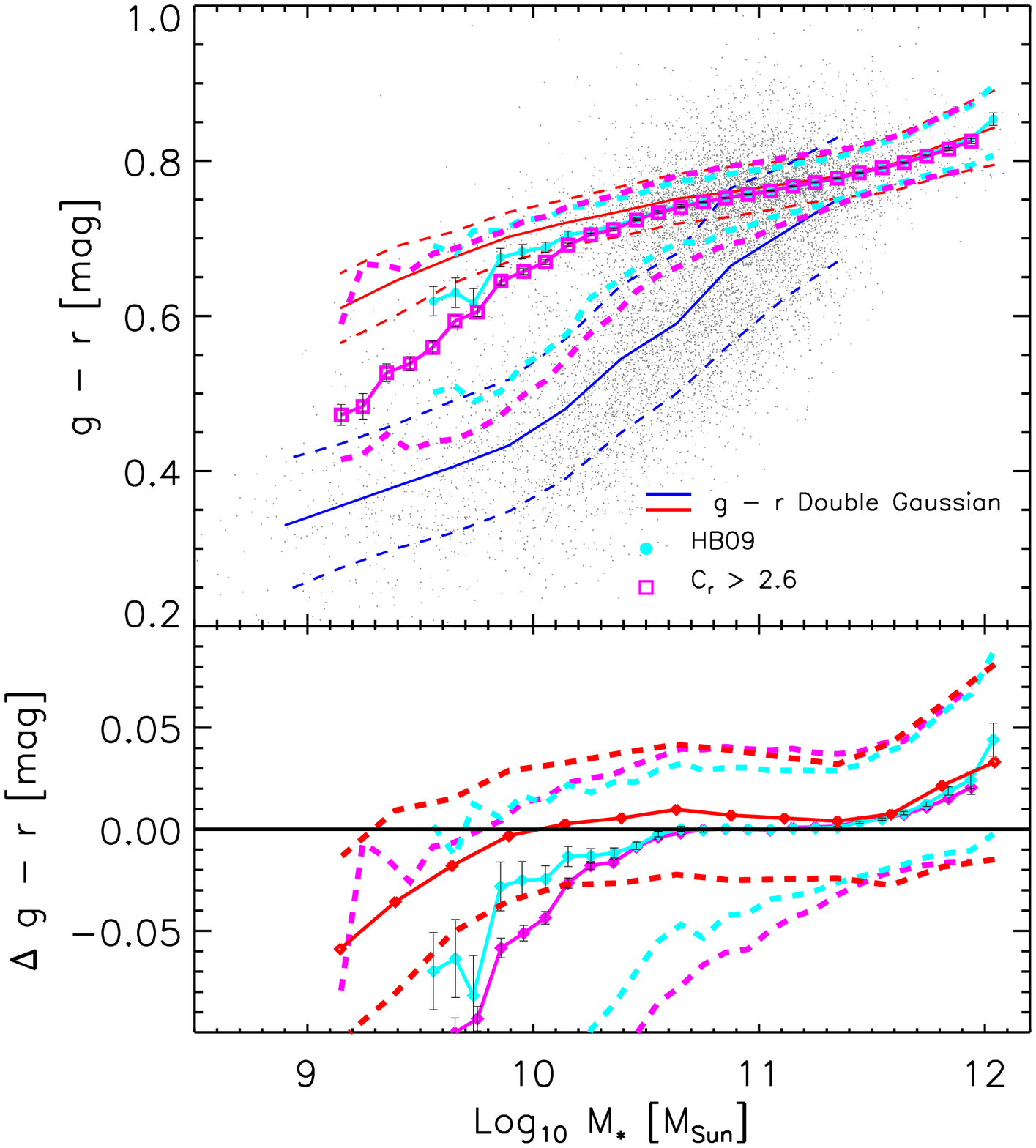}
 \caption{Dependence of the red sequence on how it is defined. 
   Panel on the left shows the color-magnitude relation; 
   panel on the right shows the color-$M_*$ relation.  
   To highlight the curvature in the upper panels, bottom panels show 
   the result of removing a linear trend: 
   $g-r - (0.361 - 0.019M_r)$ vs $M_r$ and 
   $g-r - (0.224 + 0.049\log_{10}M_*/M_\odot)$ vs $M_*$.  
   Small dots show a representative subsample of the galaxies when 
   {\tt model} colors and {\tt cmodel} magnitudes are used and the 
   $k$-correction is from Blanton \& Roweis (2007).
   Solid red and blue curves show the result of our double-Gaussian 
   decomposition (see Table~\ref{gmrMredblue} in Appendix~\ref{2gs}); 
   filled cyan circles show the color-magnitude relation for a sample 
   selected following Hyde \& Bernardi (2009); open magenta squares 
   show this relation for objects with $C_r>2.6$.  
   Dashed lines show the rms scatter around the mean relations.  
   The relation found by the double-Gaussian fit is narrower and 
   almost independent of $M_r$; the sample with $C_r>2.6$ has the 
   largest scatter, particularly at $M_r>-20$.  
 }
 \label{gmrM}
\end{figure*}

The flattening of the slope as one moves brightwards from the 
faintest luminosities is in excellent agreement with that reported 
by Skelton et al. (2009) who selected galaxies with $C_r > 2.6$ and 
at $z<0.06$.  The dot-dashed lines show the relations they reported.  
Note, however, that they did not report any upward curvature at the 
brightest end.  This may be because their sample was restricted to 
small redshifts ($z<0.06$), so they had many fewer objects at $M_r<-23$.  
As a result, at the bright end, our measurements zig-zag around their 
relation.  

While our primary interest is in the fact that the relation is curved, 
notice that the samples do have different amounts of scatter around 
the mean relation: whereas they have similar red envelopes, the 
scatter bluewards tends to increase dramatically at faint luminosities, 
with the effect being most pronounced in the $C_r>2.6$ sample.  
Some of this is because, at fainter luminosities, these samples are 
increasingly contaminated by later-type galaxies (Bernardi et al. 2010a), 
so one might worry that the steeper slope at the faint end is due, 
at least in part, to this contamination.  
In Appendix~\ref{fukugita}, we present a direct measurement of the 
color-magnitude relation in (substantially smaller) subsamples of 
fixed morphological type.  This shows that there are three distinct 
regimes in a sample composed only of Es.

\begin{table}
\caption[]{Coefficients of linear fits to the $\langle g-r|M_r\rangle$ 
  and $\langle g-r|M_*\rangle$ relations in the Hyde \& Bernardi (2009) 
 sample, where $g-r$ was computed using the Blanton \& Roweis (2007) 
 $k$-correction.}
\begin{tabular}{ccc}
\hline 
 & PETROSIAN  &\\
 Range & slope & z.p. \\
\hline 
 $-20.25 <$M$_r < -19$ & $ -0.029 \pm   0.003$ & $  0.131 \pm   0.051$ \\
 $-22 <$M$_r < -20.75$ & $ -0.014 \pm   0.001$ & $  0.434 \pm   0.012$ \\
$-23.5 <$M$_r < -22.25$ &$ -0.039 \pm   0.003$ & $ -0.104 \pm   0.061$ \\
\hline 
$9.8 < {\rm Log}_{10} {\rm M}_* < 10.2$ & $ 0.092 \pm   0.012$ & $ -0.249 \pm   0.072$ \\
$10.5 < {\rm Log}_{10} {\rm M}_* < 11.1$ & $  0.040 \pm   0.003$ & $  0.303 \pm   0.028$ \\
$11.4 < {\rm Log}_{10} {\rm M}_* < 11.9$ & $   0.094 \pm   0.002$ & $ -0.305 \pm   0.027$ \\
 \hline
 & MODEL  &\\
 Range & slope & z.p. \\
\hline 
 $-20.25 <$M$_r < -19$ & $ -0.029 \pm   0.002$ & $  0.144 \pm   0.024$ \\
 $-22 <$M$_r < -20.75$ & $ -0.019 \pm   0.001$ & $  0.361 \pm   0.013$ \\
$-23.5 <$M$_r < -22.25$ &$ -0.029 \pm   0.001$ & $  0.119 \pm   0.027$ \\
\hline 
$9.8 < {\rm Log}_{10} {\rm M}_* < 10.2$ & $  0.090 \pm   0.009$ & $ -0.215 \pm   0.093$ \\
$10.5 < {\rm Log}_{10} {\rm M}_* < 11.1$ & $  0.049 \pm   0.001$ & $  0.224 \pm   0.021$ \\
$11.4 < {\rm Log}_{10} {\rm M}_* < 11.9$ & $   0.088 \pm   0.006$ & $ -0.221 \pm   0.065$ \\
\hline \\
\end{tabular}
\label{gmrMtable} 
\end{table}

The left hand panel of Figure~\ref{gmrEtypes} shows that the 
curvature does not depend on precisely how the colors and magnitudes 
were defined:  using {\tt Petrosian} rather than {\tt model} colors 
and magnitudes makes little difference.  Appendix~\ref{systematics} 
shows that the small differences between {\tt model} and {\tt Petrosian} 
based quantities arise because {\tt model} colors probe smaller scales 
than do {\tt Petrosian} colors, and early-type galaxies have color 
gradients.  It also shows that the appearance of three regimes is 
robust against changes in the $k+e$ corrections.  

Since {\tt Petrosian} colors probe more of the total light, we will 
use them, primarily, in what follows.  
The right hand panel of Figure~\ref{gmrEtypes} shows that the three 
regimes are even more pronounced if one replaces luminosity with 
stellar mass.  
This is easily understood:  $\log M_*$ is obtained from $\log L$ by 
adding $1.097\,(g-r) - 0.406$.  So, to make this plot, one slides the 
reddest objects in the previous plot to the right, and the bluest 
to the left.  
Table~\ref{gmrMtable} shows that the slope at intermediate masses is 
a factor of two shallower than at either end.  The changes in slope 
occur at $M_*=3\times 10^{10}M_\odot$ and $M_*=2\times 10^{11}M_\odot$.  

\begin{figure}
 \centering
 \includegraphics[width=0.95\hsize]{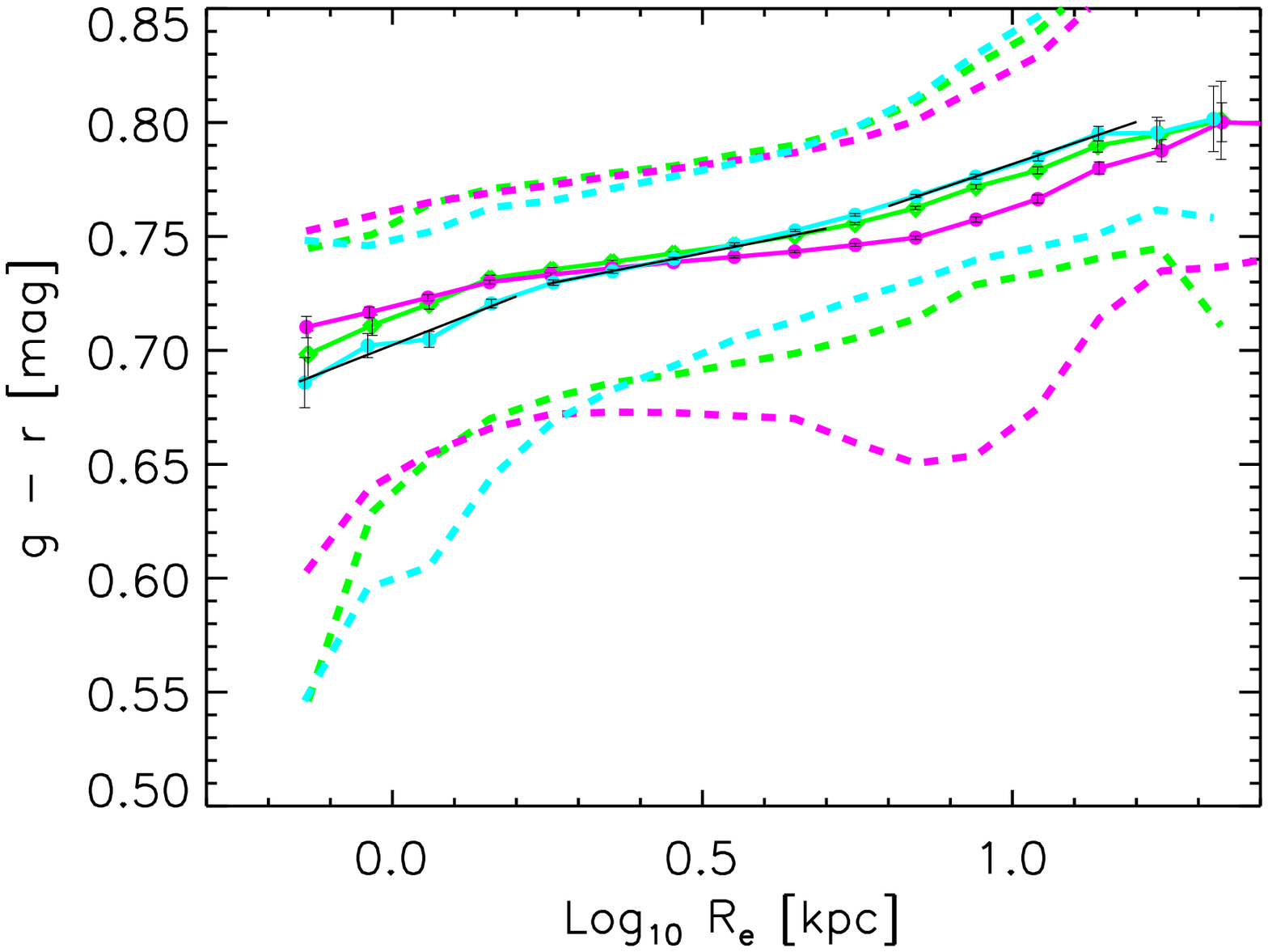}
 \includegraphics[width=0.95\hsize]{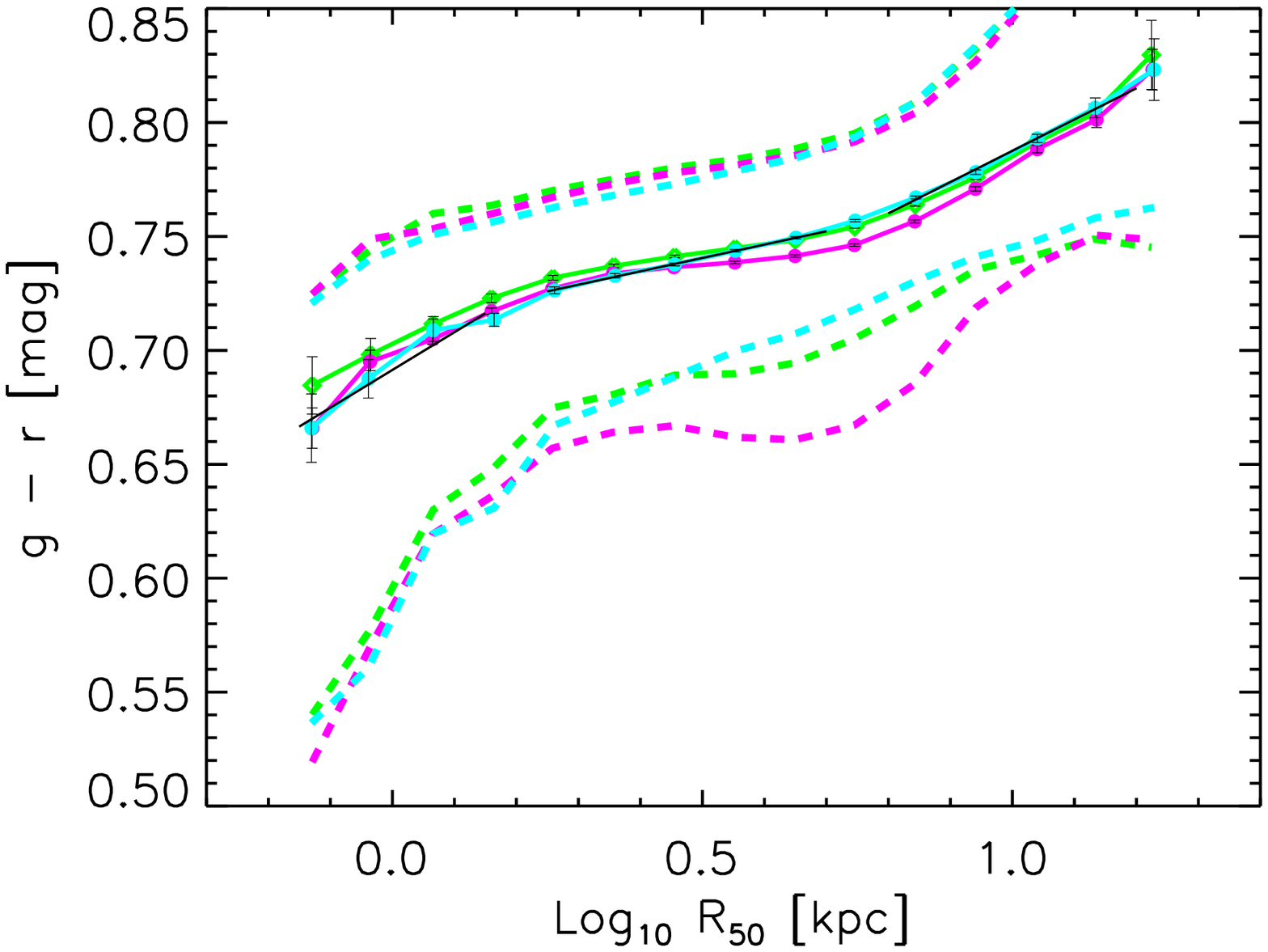}
 \caption{{\tt Petrosian} color vs size, {\tt cmodel} $R_e$ (top) and 
          Petrosian ${\tt R_{50}}$ (bottom), for the three ways of selecting 
          early-type samples (compare Figure~\ref{gmrEtypes}).  
          }  
 \label{gmrSize}
\end{figure}

Figure~\ref{gmrM} displays the red and blue sequences defined by 
our double-Gaussian fits described in Appendix~\ref{2gs} (the 
parameters are reported in Table~\ref{gmrMredblue}).  Clearly, 
the red sequence defined in this way also shows three regimes.  
Notice that the red sequence is considerably straighter and narrower 
than the blue, and that the thickness of the two sequences is almost 
independent of luminosity, even though this was not required during 
the fitting procedure.  
This is significant, because we were previously concerned that the 
bluewards flaring in the other samples might be signalling that the 
mean relation had been affected.  Here, that argument cannot be made.  
Nevertheless, the mean red sequence is curved, in excellent agreement 
with that shown in Figure~\ref{gmrEtypes}.  
The panel on the right shows that the three regimes are also present 
if one replaces luminosity with stellar mass.  Table~\ref{gmrMsredblue} 
provides details of the double-Gaussian fits to the associated red and 
blue sequences.  

Before moving on, it is worth noting that the double-Gaussian fits 
do not fare well over the range $-18.5 \ge M_r\ge -20.5$ 
(see Figure~\ref{histgmr}).  At these luminosities, there appears 
to be a set of objects which populate the `green valley' between 
the red and blue sequences.  However, this third component is most 
needed at luminosities which lie below those where the color-magnitude 
relation flattens.  So our neglect of, or contamination by this 
component is not to blame for the flattening at intermediate 
luminosities, nor for the steepening at the highest luminosities.  

\begin{figure}
 \centering
 \includegraphics[width=0.95\hsize]{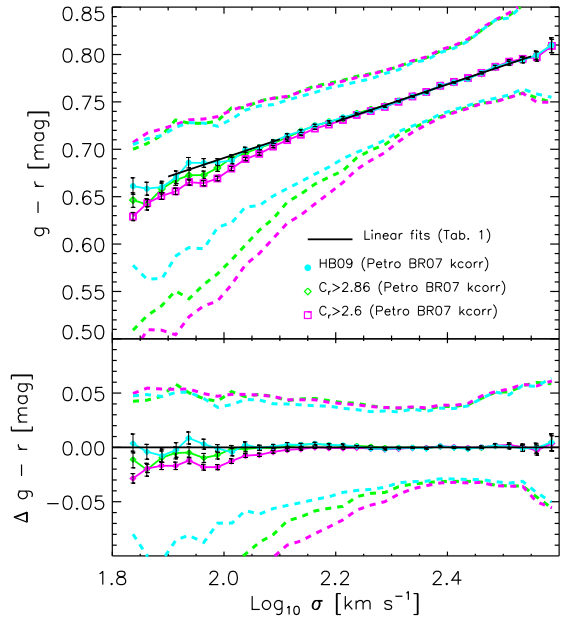}
 \caption{{\tt Petrosian} color vs velocity dispersion 
          for the three ways of selecting early-type samples 
          (compare Figures~\ref{gmrEtypes} and \ref{gmrSize}).  }
 \label{gmrSigma}
\end{figure}



\subsection{Curvature in the color-$R$ and $M_{\rm dyn}$ relations but little in color-$\sigma$}

In contrast to the previous two correlations with color, the 
color-size relation has been much less studied.  
Figure~\ref{gmrSize} shows that the color-$R$ relation also shows 
three distinct regimes.  These are somewhat more obvious if we use 
{\tt Petrosian} ${\tt R_{50}}$ than {\tt cmodel} $R_e$.   
Three distinct regimes are also seen if the dynamical mass is used 
instead of stellar mass, although the curvature at the high-mass 
end is less steep (we have chosen to not show this plot).  

In contrast to these relations which are rather curved, the 
color-$\sigma$ relation is rather well described by a single power law.  
This is shown in Figure~\ref{gmrSigma}.  At large $\sigma$, the 
relation is independent of how the sample was selected.  However, 
at $\log_{10}(\sigma/{\rm km~s}^{-1}) < 2.1$, samples which are more 
likely to include later types fall below the relation, suggesting 
that it is the changing morphological mix which is driving the 
curvature at small $\sigma$.  
Since major mergers are expected to change the mass and size of a 
galaxy while leaving $\sigma$ unchanged, the lack of curvature at 
large $\sigma$ is suggestive.  
We return to this in Sections~\ref{sec|theory} and~\ref{discuss}. 

Before moving on, we note that, for the bulk of the early-type population, 
the color-magnitude relation is a consequence of the color$-\sigma$ and 
$\sigma-$magnitude relations (Bernardi et al. 2005).  
This means that $\sigma$ determines both the color and the luminosity of 
an object, at least for the bulk of the population at lower luminosities.  
Now, the $\sigma-$magnitude relation flattens at large luminosities 
(Bernardi et al. 2007), and there is no curvature in the color-$\sigma$ 
relation (Figure~\ref{gmrSigma}).  Hence, if there were no scatter 
around these relations, we would expect the color-magnitude relation 
to flatten rather than steepen at $M_r<-22.5$.  Therefore, either 
$\sigma$ is no longer the important parameter at these high luminosities 
(and stellar masses), or the scatter around these relations is important.


\section{Dependence on age and metallicity of the population}\label{sec:ageZ}
The previous subsections showed that the curvature in the 
color-magnitude relation is clearly seen in a number of other 
scaling relations with luminosity or stellar mass, but is essentially 
absent in correlations with $\sigma$.  
We now turn to a study of how the curvature depends on the age and 
metallicity of the population.  Gallazzi et al. (2006) have shown 
that both age and metallicity increase along the color-$M_*$ 
relation.  Here, our primary interest is in seeing if the curvature 
we have found is associated with stellar population effects.  

\begin{figure}
 \centering
\includegraphics[width=0.95\hsize]{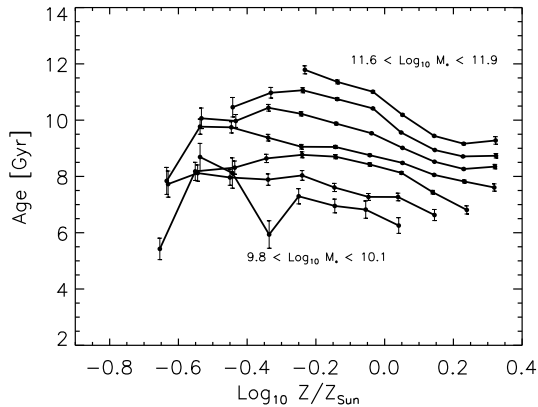} 
 \caption{Age-metallicity relation for a number of bins in stellar mass.  
  Although more massive objects are older and more metal rich, at 
  fixed mass, older objects are more metal poor.}
\label{fig|AgeMetMs}
\end{figure}

Our age and metallicity estimates come from Gallazzi et al. (2005); 
they are based on absorption line features in the spectra.  
About 50 percent of our sample has ages between 8 and 10 Gyrs; 
20 percent have ages between 10 and 12 Gyrs and only a percent 
are older than 12 Gyrs; 
20 percent have ages between 6 and 8 Gyrs, and about 7 percent are younger 
than 6 Gyrs.
Figure~\ref{fig|AgeMetMs} shows that, although both age and 
metallicity tend to increase with mass, at fixed $M_*$, age and 
metallicity are anti-correlated:  
older galaxies are more metal poor, in agreement with previous work 
(Trager et al. 2000; Bernardi et al. 2005).

\begin{figure}
 \centering
\includegraphics[width=0.95\hsize]{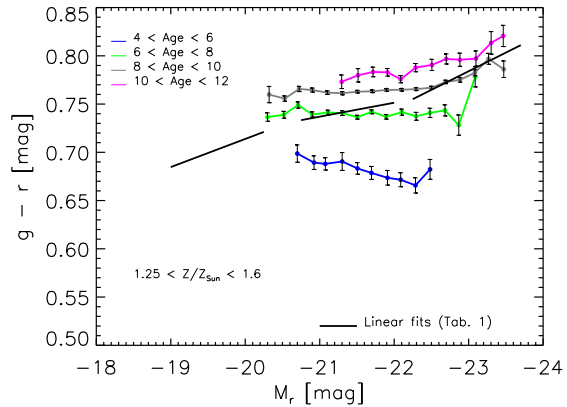} 
\includegraphics[width=0.95\hsize]{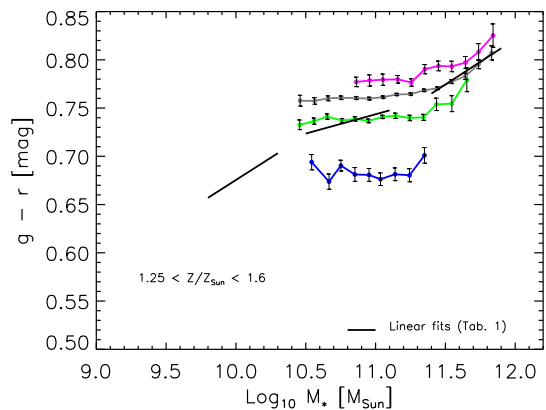} 
\includegraphics[width=0.95\hsize]{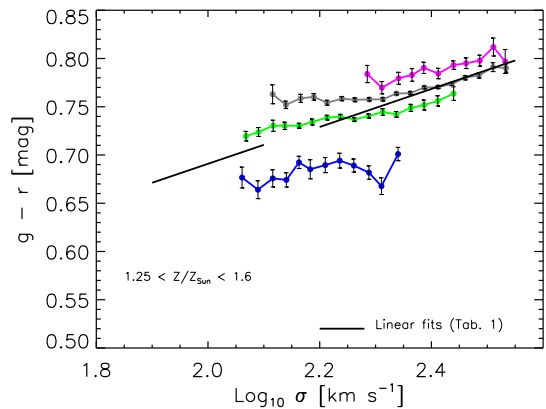} 
 \caption{Color-magnitude, $M_*$ and $\sigma$ relations for galaxies 
 with fixed age and metallicity (as indicated).  Curvature in the 
 top two panels is seen only for the oldest, most massive objects.}
\label{fig|AgeMet}
\end{figure}

The top panel of Figure~\ref{fig|AgeMet} shows that, at fixed metallicity 
and age, the color-magnitude relation is flat for galaxies with 
$M_r>-22.5$.  The Figure actually shows results for metallicities between 
$1.25-1.6 Z_\odot$.  At smaller metallicity (not shown), the colors 
for the same age bins are offset blueward with respect to those 
shown here, but the color-magnitude relation remains flat.  
(This is because colors suffer from an age-metallicity degeneracy; 
Gallazzi et al. used spectral line indices to break this degeneracy.) 
In fact, the relation is flat whatever the age or metallicity.  
The middle panel shows this is true for the color-$M_*$ relation 
(at $\log_{10} M_*/M_\odot < 11$) as well.  

However, the color increases with luminosity (top) and even more 
strongly with $M_*$ (middle), at the most massive end which is 
dominated by the oldest galaxies.  For younger galaxies, the upturn 
may be due to correlated errors in the $M_*$ and age estimates, but 
this is not a concern for the older objects (see Bernardi 2009 for 
more discussion).  

This upwards curvature is not seen in the color-$\sigma$ relation (bottom).  
The slight increase of $g-r$ with $\sigma$, at fixed age and metallicity, 
may be due in part to the fact that the model estimates assume that all 
galaxies have the same ratio of $\alpha$-elements with respect to Fe.  
(Models which account for variations in $\alpha$-abundance are only 
just becoming available -- they were not available to Gallazzi et al.) 
However, this ratio is known to be strongly correlated with $\sigma$:  
large $\sigma$ galaxies are $\alpha$-enhanced (Trager et al. 2000; 
Bernardi et al. 2005).  Thus, while it may be that the results shown 
in the bottom panel are biased because this correlation has been ignored, 
it is extremely unlikely that the upwards curvature in the other two 
correlations (at fixed age and metallicity) is due to $\alpha$-enhancement 
related biases.  

It is interesting that this same age and stellar mass scale is seen 
in recent studies of the $R_e-M_*$ relation.  Shankar \& Bernardi (2009) 
show that, at $M_*<2\times 10^{11}M_\odot$, older early-types tend to 
have smaller sizes than younger ones, perhaps because they formed 
at higher redshift from more dissipative mergers.  However, at 
higher masses, this dependence on formation time disappears.  
Shankar \& Bernardi suggest that this is because some later process 
has erased the trend.  Although it is possible that some process 
decreases the sizes of younger early-types, Shankar \& Bernardi 
focus on the possibility that the sizes of the older ones have 
increased (see also Shankar et al. 2010a).  They argue that if older 
objects have undergone more dry mergers than their younger 
counterparts (of the same mass), then this would puff up their sizes, 
effectively erasing the trend which derives from formation age/time.  

\section{Dry merger models}
\label{sec|theory}

In this section we are particularly interested in assessing if a late,
dry merger-driven evolution for massive and passive early-type galaxies,
is consistent with the measurements presented earlier in this paper.

Semi-analytic galaxy formation models make predictions for the curvature
and evolution of the color-magnitude relation, so, in principle, they
could be used to address this question.  However, Bernardi et al. (2007)
have shown that the red-sequence in these models is too red, and although
it turns blueward at intermediate luminosities, it does not turn redward
at the highest luminosities.
In addition, Shankar et al. (2010a,b) have shown that the bulge sizes in
some models could be somewhat discrepant with measurements.
 
Therefore, we now discuss a number of plausible scenarios in light 
of our measurements, some of which we simulate numerically.  
These are toy models:  they do not provide a precise quantification 
of the color evolution.  In what 
follows, we will assume that after some sufficiently large redshift, 
which we will take to be $z\sim 1$, the stars evolve passively, and 
this evolution is not differential.  The absence of differential 
evolution means that we can effectively remove its effects from the 
following discussion, as including it simply results in an overall 
translation of the objects in the color-magnitude plane, but does 
not alter any features in the color-magnitude relation.  
To the extent that differential evolution is expected, 
it goes contrary to the trend that we observe:  the most massive 
objects are expected to contain the oldest stars, so their luminosities 
and colors are expected to evolve more slowly than those of the 
least massive objects.  Hence, while differential evolution may 
contribute to the flattening of the color-magnitude relation at 
intermediate luminosities, it seems an unlikely explanation for the 
steepening towards redder colors at large $M_*$.  

Finally, we note that all the models we describe below assume that 
objects which are on the blue sequence at $z=1$, but evolve on to 
the red sequence as their gas supply is removed or exhausted -- i.e., 
no mergers are involved -- are either a negligible fraction of the 
population or, when they join the red sequence, they do so with colors 
that are representative of the red population for their mass (e.g., 
they are not biased bluewards), and they then evolve via dry mergers 
similarly to the other objects that were already on the red sequence.  
This assumption is consistent with the recent results of Peng et al. (2010)
(see their Fig. 13 and 16) and Eliche-Moral et al. (2010) (who suggest 
that $z\sim 0.8$ might be more appropriate).  

\subsection{Similar initial conditions, but two types of merger histories (Model I)}
Suppose that at some sufficiently high redshift (which we will take to be 
$z\sim 1$), the color magnitude relation was approximately a power law, 
and that, thereafter, the stars evolve passively, and this evolution is 
not differential.  Then, dry mergers will cause the color magnitude 
relation to curve bluewards (from the initial power law) at the 
bright end, with the amount of curvature depending on the typical mass 
ratio of the mergers, and how that ratio depends on mass.  We will 
loosely refer to mass ratios of 0.3:1 or greater as being major mergers, 
and smaller ratios as being minor.  

\begin{figure}
 \centering
\includegraphics[width=0.95\hsize]{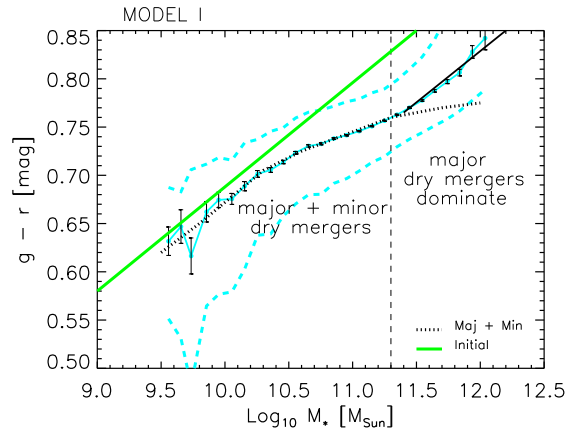} 
 \caption{Origin of the color-$M_*$ relation if the stellar mass growth 
   at $\log_{10}M_*/M_\odot > 11.3$ is through major (1:1) dry mergers only.  
   Such mergers translate the initial power-law relation (green solid line, here 
   assumed to be for $z=1$) to the right (black solid line offset to right).  
   Minor mergers at lower masses tend to flatten the relation from the 
   initial power-law (dotted line), because the merger product must be 
   bluer than its most massive progenitor.  Given an initial relation 
   (here assumed to be a power-law at $z=1$) the free parameters in this 
   model are the mass scale at which 1:1 mergers dominate, the ratio 
   of minor to major mergers at lower masses.  Solid cyan line and 
   associated error bars show the relation we see at $z\sim 0$; 
   dashed lines show the rms spread around this mean relation.}
\label{fig|Mssigma1}
\end{figure}

Suppose that objects which are low mass today were assembled from both 
minor and major mergers, whereas the most massive objects experienced 
only 1:1 mergers.  Then, the color magnitude relation will be flattened 
from the initial power law at low luminosities (minor mergers make the 
merged product bluer), but it will simply be translated to the right 
at high luminosities. Figure~\ref{fig|Mssigma1} shows this schematically.
By adjusting the total mass growth and ratios at low masses, and the 
mass scale at which the mergers become major only, this scheme can be 
made to agree with our measurements.  

In this model, the lack of curvature in the color-$\sigma$ relation 
can be understood as follows.  The major 1:1 mergers will change 
neither $g-r$ nor $\sigma$, so they still lie on the initial 
relation.  Minor mergers which decrease the color also decrease 
$\sigma$; this partially removes the flattening (in color-$\sigma$) 
which is so much more evident in the color-magnitude relation.  Thus, 
in this model, the color-$\sigma$ relation at $z\sim 0$ differs from 
that at $z\sim 1$ primarily because of passive evolution of the 
colors -- if the evolution is not differential, then the local 
relation is simply offset from that at higher $z$.  
In addition, whereas major mergers change the size proportionally to 
the mass, minor mergers change the sizes more than the masses.  This 
accounts for the larger range in $R_e$ for which the color-$R_e$ slope 
is shallow.  

It is worth stating explicitly that this model works because there 
is a color-magnitude relation at $z\sim 1$.  Then, the additional 
requirement that the most massive galaxies are formed from major mergers, 
means that the most massive galaxies today formed from objects that 
were redder than those which make intermediate mass galaxies.  
Bernardi et al. (2007) noted that just such a conspiracy of 
mass/color-dependent mergers was required to explain the red colors
of BCGs.  Unfortunately, there is no obvious choice for the transition 
mass scale which plays a crucial role in this model, although, as we 
now argue, color gradients may hold an important clue.  

In particular, Roche et al. (2010) show that color gradients are maximal 
at $M_r=-22$ (see our Figure~\ref{gmrMPM}).  Whereas major mergers are 
expected to decrease color gradients, minor mergers should not change 
the gradients significantly, or they may enhance them slightly.  
This is because the smaller bluer object involved in the minor merger 
will deposit most of its stars at larger distances from the center of 
the object onto which it merged.  Bernardi et al. (2010b) show that 
this same scale appears in other scaling relations as well.  
Thus, it may be that $M_r<-22$, which corresponds to 
$M_* > 2\times 10^{11}M_\odot$, is the scale above which major mergers 
dominate.  

\begin{figure}
 \centering
\includegraphics[width=0.95\hsize]{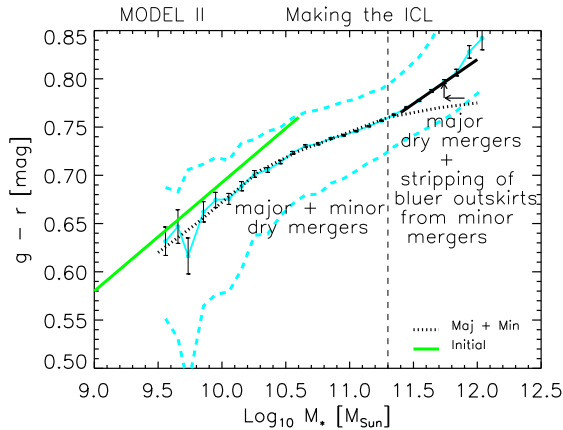} 
 \caption{Origin of the color-$M_*$ relation if the stellar mass growth 
   at the most massive end is through major dry mergers, because minor 
   mergers (in this case, at $\log_{10} M/M_\odot > 11.3$) contribute 
   primarily to the ICL.  
   Note that although these minor mergers will not change the stellar 
   mass, they will affect the size and velocity dispersion (hence 
   dynamical mass) of the merger product.  
   Green solid line shows the relation we assume at $z=1$, dotted line shows 
   the relation at $z=0$ due to minor mergers at lower masses, black solid line 
   shows the $z=0$ relation at large $M_*$, where the minor mergers 
   contribute to the ICL rather than to $M_*$, and solid cyan line with 
   associated error bars and dashed lines (same as previous figure) show 
   our $z\sim 0$ measurements.}
\label{fig|Mssigma2}
\end{figure}

\subsection{Similar initial conditions, but inclusion of stripping/ICL (Model II)}\label{model2}
This model is similar to the previous one, except that we assume that 
the massive end is dominated by objects in clusters, for which the 
effects of tidal stripping etc. matter (Figure~\ref{fig|Mssigma2}). 
In this case, we assume that mergers at the high mass end may be both 
major and minor, but that sufficiently minor mergers do not actually 
contribute to the stellar mass of the final object, because they will 
be shredded; they contribute to the intercluster light.  
Note that minor mergers will produce changes in the size and velocity 
dispersion (hence dynamical mass) of the merger product, just not to
the stellar mass.  Thus, although the assembly history of BCG-like 
objects will involve both minor and major mergers, the stellar mass 
only grows in major mergers.  

The net result will be similar to the previous model, with shallowing 
at low masses where both types of mergers happen (but stripping does 
not), and a parallel shift to larger masses of the initial (steeper) 
relation at the high luminosity end (where stripping erases the effects 
of minor mergers on the mass growth).  I.e., in this model, the 
transition mass scale is related to the formation of the ICL.  

Note that color gradients of the satellites which are stripped means 
that stars which do make it all the way to the central object will be 
redder, further steepening (or producing less flattenning of) the 
color-$M_*$ relation at the massive end.  (If so, the ICL should be 
bluer than the BCG.)  In addition, because some mass is lost to the ICL 
(some estimate that there is at least as much mass in the ICL as  
there is in the BCG), the color-magnitude relation will not extend to 
as high masses as in our first model.  And finally, in this model, the 
ratio of stellar to dynamical mass should decrease at large masses, 
in qualitative agreement with the observation that 
$M_* \propto M_{\rm dyn}^{0.75}$.  On the other hand, by consigning 
to the ICL some of the stellar mass that would otherwise have ended up 
in massive objects, this model is constrained by recent work suggesting 
that there is 50\% more mass in objects with $\log_{10}M_*/M_\odot > 11.3$ 
than previously thought (Bernardi et al. 2010b).  This model must make 
such objects, as well as the ICL.  

\subsection{Correlation between color-magnitude residuals and mergers (Model III)}
\begin{figure}
 \centering
\includegraphics[width=0.95\hsize]{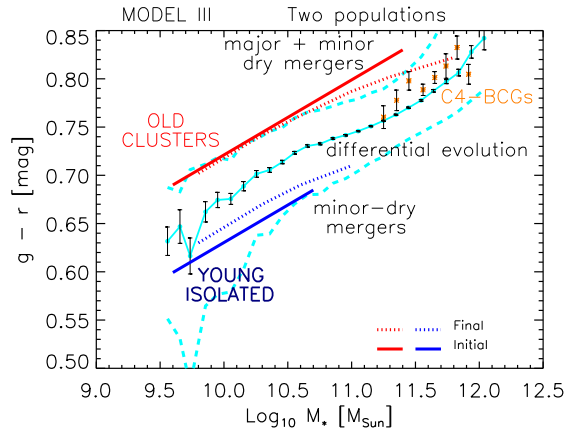} 
 \caption{Origin of color-$M_*$ relation if the oldest objects at the 
  present time formed from the oldest, reddest objects in the past, 
  through a sequence of major and minor dry mergers, whereas the 
  youngest objects today formed from minor mergers of bluer objects. 
  Upper red solid line shows the assumed color-$M_*$ relation of the oldest 
  objects at $z=1$; lower blue solid line shows that for the youngest objects 
  at $z=1$.  The associated dotted lines show how these relations have 
  changed by $z=0$.  Error bars and associated dashed lines (same as 
  in previous Figure) show $\langle{\rm color}|M_*\rangle$ for the 
  full sample at $z\sim 0$, and orange filled circles with error bars 
  show this relation for BCGs at $z\sim 0$.}
\label{fig|Mssigma3}
\end{figure}

In this model, we distinguish between objects which lie redward of the 
mean color-magnitude relation at $z=1$, and those which lie blueward
(Figure~\ref{fig|Mssigma3}). 
Here, we assume that the redder objects are older, in agreement with the 
trend at $z=0$ (Kodama et al. 1998; Bernardi et al. 2005).  We then 
assume that these redder objects were involved in major and minor 
mergers, whereas the bluer objects only experienced the most minor 
mergers, so they have increased their mass little since $z\sim 1$.  

If our previous model of stripping which contributes to the formation of 
ICL (i.e. Model II) is realistic, then, in the present context, it would 
apply only to the redder objects.  However, by ensuring that red objects 
merge with red ones, mergers in this model produce less of a decrease in 
slope, so less is required of processes like stripping to reproduce the
turn up towards redder color at the high-mass (luminosity) end.  
Therefore, it may be easier for this model to produce the observed 
amount of stellar mass locked-up in objects with 
$\log_{10}M_*/M_\odot > 11.2$ at $z\sim 0$.  

In many respects, this model is a variant of Model I.  
There, major mergers were used to ensure that the most massive objects 
formed by mergers of the reddest objects.  The present model achieves 
this by assuming that massive objects are older, rather than making 
a specific assumption of major vs minor mergers.  Additionally, here, 
the minor mergers which produce the lowest mass galaxies are 
preferentially of bluer objects, so they tend to result in bluer colors 
today.  Thus, the conspiracy of mass/color-dependent mergers noted by 
Bernardi et al. (2007) to explain BCGs is here extended to the faint end 
as well.  Note that, at the high mass end, this conspiracy may alleviate 
the tension between $\alpha$-enhancement ratios and late assembly models 
that has been emphasized by Pipino \& Matteucci (2008).  

In the previous models (i.e. Model~I and II), the curvature is 
determined by the slope of the $z\sim 1$ color-magnitude relation:  
a flatter slope produces a smaller effect.  Here, the scatter in the 
$z=1$ color-magnitude relation also matters:  a smaller scatter also 
produces a smaller effect (e.g., if there were no scatter around the 
$z\sim 1$ relation, we would have no Model~III).  

It may help to think of the cluster population at $z=0$ as being made 
of these redder objects, whereas the bluer objects are now in lower 
density regions.  
This raises a potential problem because, at the present time, the 
environmental dependence of the color-magnitude relation of early-type 
galaxies is small (e.g. Bernardi et al. 2006).  
However, two effects in this model serve to help meet this constraint.  
First, the flattening and rightwards shift of the sequence defined by 
the older galaxies (red dotted line in Figure~\ref{fig|Mssigma3}) brings 
it into better agreement with the relation defined by extrapolating the 
relation of the younger objects (blue solid line) to higher masses, 
thus reducing the offset in colors between cluster and field galaxies 
that would otherwise result.  And second, differential evolution 
(because now we explicitly assume the populations have different ages) 
also acts to erase the offset in colors between the younger and older 
galaxies (which we have schematically represented by shifting the 
dotted blue line slightly redwards of the solid blue line).  
Together, both effects also make the scatter in the color-magnitude 
relation smaller at $z=0$ than at $z=1$.  Note that, in addition to 
this testable prediction, this model also suggests that the residuals 
from the high redshift relation should correlate more strongly with 
environment than they do today.  This is because, in this model, objects 
which are redder than average at $z\sim 1$ are in clusters at $z=0$ -- 
but in hierarchical structure formation models, objects in clusters today 
were in overdense regions in the past (e.g. Mo \& White 1996; Sheth 1998; 
Sheth et al. 2006).

\begin{figure}
 \centering
 \includegraphics[width=\hsize]{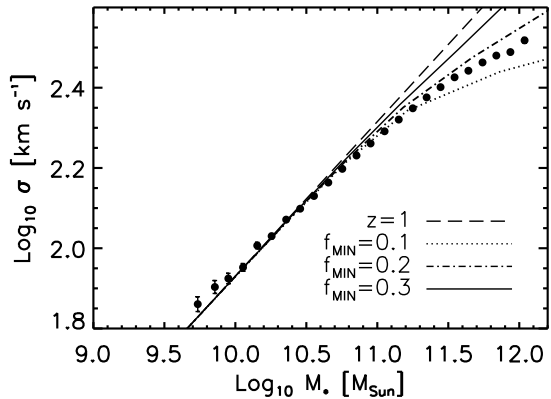}
 \caption{Stellar mass versus velocity dispersion in the SDSS
 (filled symbols).  Long-dashed line shows our assumed relation
 at $z=1$.  The other curves show the result of evolving it down
 to redshift $z=0$, as detailed in the text, for three choices
 of $f_{\rm MIN}$.}
\label{fig|Mssigma}
\end{figure}

\subsection{Numerical implementation of Model I}
To illustrate Model I, we have performed crude numerical simulations in 
which we prescribe the joint distribution of color, stellar mass and 
velocity dispersion at $z=1$.
We then let these galaxies merge at the rate expected from
observations and halo occupation modelling, always assuming zero-energy 
orbits with no energy dissipation.  The assumption that both the 
initial objects and the final ones are in virial equilibrium allows 
one to determine the scaling relations of the population at late 
times from those of the initial population (see Appendix~\ref{0energy} 
for details).  We then compare the resulting scaling relations with 
our measurements in the SDSS at $z=0$.
Note that this approach assumes that objects which are on the blue 
sequence at $z=1$, but evolve on to the red sequence as their gas 
supply is removed or exhausted -- i.e., no mergers are involved -- 
are a negligible fraction of the population. 

We set the scaling relations of the initial population as follows.  
We assume the color-$M_*$ relation has the same slope at $z=1$ as at 
$z=0$; this is consistent with observations (Mei et al. 2009).
We then assume that the $\sigma-M_*$ relation at $z=1$ is the same
power-law (both slope and zero-point) as the faint end of the $z=0$
relation.  At $z=0$ the color-$M_*$ slope equals the product of the
color-$\sigma$ and $\sigma-M_*$ slopes (Bernardi et al. 2005); we assume
this is also true at $z=1$.  So the one free parameter is the zero-point
of the $z=1$ color-$M_*$ relation; setting it also determines the zero-point
of the $z=1$ color-$\sigma$ relation.
The low-$z$ dependence of the color-magnitude relation suggests that
the $g-r$ colors are bluer at higher redshifts by approximately $0.2z$
(Figure~\ref{gmrEvolM}).  Therefore, we assume that they continue to
evolve in this way upto $z=1$.


\begin{figure}
 \centering
 \includegraphics[width=\hsize]{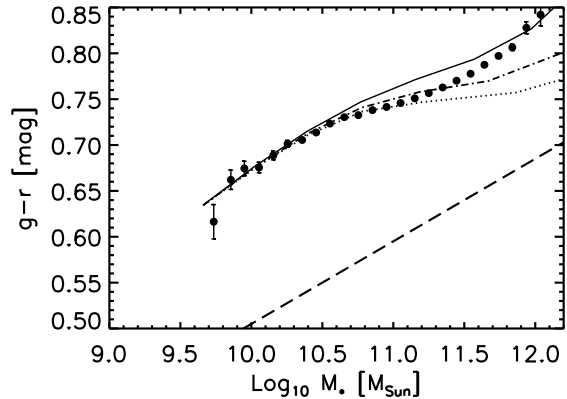}
 \includegraphics[width=\hsize]{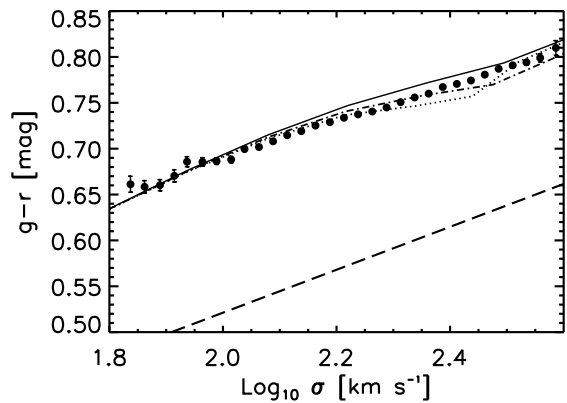}
 \caption{Same as Figure~\ref{fig|Mssigma} but for the
 color-$M_*$ (top) and color-$\sigma$ (bottom) relations.}
 \label{fig|ColMS}
\end{figure}

We then evolve the $z=1$ relations down to $z=0$ by a sequence of 
dry mergers.  We do so by dividing the interval $0<z<1$ into ten 
discrete steps.  For each, we estimate how the dry merger rate 
depends on stellar mass following Hopkins et al. (2010).  
This uses a convolution of the host halo merger rates with the 
average stellar-to-halo mass relation at each redshift, while also 
taking into account the gas fraction involved in each merger event 
(see Hopkins et al. 2010 for details, who provide a numerical 
algorithm to implement their model).
The relevant merger rates are in broad agreement with a variety of
direct observations (e.g Hopkins et al. 2010; Robaina et al. 2010 
and references therein) and other theoretical estimates 
(e.g., Guo \& White 2008).
For consistency with the observed rather passive evolution characterizing
the bulk of early-type galaxies (e.g Wake et al. 2008), we only consider
dry mergers (with $f_{\rm gas} \le 0.1$) as drivers of the late-time
evolution.  The exact threshold for $f_{\rm gas}$ does not change the 
overall trends discussed below.

The most important feature of these merger rates is that the
evolutionary paths of the highest stellar mass bins are characterized
by a larger number of major dry mergers.  As we show below, this means
that the colors of the objects which merge to make the most massive
galaxies today are typically redder than those which make intermediate
mass galaxies, whereas the more minor mergers which produce the lowest
mass galaxies are preferentially of bluer objects, so they tend to
result in bluer colors today.  This is precisely the conspiracy of 
mass/color-dependent mergers the Bernardi et al. (2007) argued was 
required to explain the red colors of BCGs -- our model quantifies 
the resulting trends.

The merger rates, and our results, depend on the mass ratio of the 
merging objects.  If $m$ is the initial object in a given time step, 
then the merged object has mass $m(1+f)$.  
An object initially of mass $m_0$ and velocity dispersion $\sigma_0$, 
which undergoes $N$ zero-energy (sometimes called parabolic) dry mergers 
with other objects of mass $m_i<m_0$ and velocity dispersion $\sigma_i$, 
will result in an object of mass $M_f$ and velocity dispersion $\sigma_f$ 
which are given by 
\begin{eqnarray}
 M_f &=& m_0\,(1+\sum_{i=1}^N f_i) \nonumber \\
 \sigma_f^2 &=& \sigma_0^2\,
 \frac{1 + \sum_{i=1}^N f_i\,(\sigma_i/\sigma_0)^2}{1 + \sum_{i=1}^N f_i}\ ,\\
 \frac{L_r}{L_g} &=& \frac{\sum_i L_{ri}}{\sum_i L_{gi}} 
  = \frac{L_{r0}}{L_{g0}}\frac{1 + \sum_i L_{ri}/L_{r0}}{1 + \sum_i L_{gi}/L_{g0}} 
    \nonumber\\
 &=& \frac{L_{r0}}{L_{g0}}\frac{1 + \sum_i f_i (\Gamma_{r0}/\Gamma_{ri})}
      {1 + \sum_i f_i (\Gamma_{r0}/\Gamma_{ri}) (L_{r0}/L_{g0})/(L_{ri}/L_{gi}) } 
     \nonumber
\label{eq|evol}
\end{eqnarray}
where $f_i \equiv m_i/m_0 < 1$ and $\Gamma_r \equiv (M_*/M_\odot)/(L_r/L_\odot)$.  
Since $(g-r)_0 > (g-r)_i$, the expression above shows that the merger 
product $M_f$ would be bluer than $m_0$ if we were to ignore the aging 
of the stellar population.  
To simplify this expression for the color further, we assume that 
$\log \Gamma_r \propto 1.097\,(g-r)$ with a redshift dependent 
zero-point.  Since our expression only involves ratios of $\Gamma_r$s, 
this zero-point cancels out, making
\begin{eqnarray}
g-r &=& (g-r)_0 + 2.5\log_{10} \left(\frac{1 + \sum_i f_i\, 10^{1.097\,\Delta c_i}}
      {1 + \sum_i f_i\, 10^{1.497 \,\Delta c_i}}\right) \nonumber\\
 \Delta c_i &\equiv& (g-r)_0 - (g-r)_i.
\label{eq|col}
\end{eqnarray}

In our analysis, we require $f> f_{\rm MIN}$, and we study how our 
results change as we increase $f_{\rm MIN}$.  
In practice, we divide the interval  $0<z<1$ into a set of ten discrete
time steps.  
For a given time bin, we pick three equally spaced bins of 
$f$ which satisfy $f_{\rm MIN}\le f \le 1$ (we show results for three 
choices of $f_{\rm MIN}$). For each bin of $f$ we first compute the 
mean number $N_m$ of expected dry mergers undergone by a galaxy of 
mass $m_0$ and velocity dispersion $\sigma_0$ with others of mass 
$m_1 = f m_0$ and velocity dispersion $\sigma_1$, and then update 
mass, size, velocity dispersion and color according to the relations 
discussed above.
After the mass-weighting update of the colors which results from the
dry merger, we shift them redwards by $0.2(z_j-z_{j+1})$, where $z_j$
denotes the redshift associated with the time-bin just before the merger.
We then iterate from $z=1$ down to 0.
Notice that $\sigma_0 < \sigma_1$ as $m_0>m_1$; if neither $f$ nor
$N$ vary with $m_0$, then, at late times, the $\sigma-M$ relation
shifts towards smaller $\sigma$ for a given $M$.  In practice, since
$fN$ increases with $M$, the downwards shift is larger for the most
massive objects.  This makes the $\sigma-M$ relation flatten at large $M$.

Figure~\ref{fig|Mssigma} shows the $M_*-\sigma$ relation for three 
choices of $f_{\rm MIN}$:  solid, dot-dashed, and dotted lines represent 
models with $f_{\rm MIN}=0.3,0.2,0.1$, respectively,
while the long-dashed line shows the assumed $z=1$ relation.
Figure~\ref{fig|ColMS} shows the associated changes to the color-$M_*$
and color-$\sigma$ relations.
Notice that our dry merger models produce strong breaks in the $z=0$ 
color-$M_*$ relation while keeping the color-$\sigma$ relation closer 
to a power-law (bottom panel of Figures~\ref{fig|ColMS}), in reasonable 
agreement with our measurements.



\section{Discussion}\label{discuss}
Our study of the color-magnitude and color-$M_*$ scaling relations 
has revealed interesting trends: one at $M_*\sim 3\times 10^{10}M_\odot$ 
which had been noticed before (Kauffmann et al. 2003; Skelton et al. 2009), 
and another at high luminosities ($M_r\le -22.5$) and masses 
($\log_{10}(M_*/M_\odot)\ge 11.3$), which is new to our work.  
These trends are qualitatively independent of exactly how the early-type 
sample is selected.  In most cases, the (weak) dependence on precisely 
how the sample was selected can be traced to contamination of the 
red-sequence by edge-on spirals.  
In a related paper, Bernardi et al. (2010b) show that a number of 
other scaling relations also indicate that these luminosity 
and mass scales are special.  

The red sequence is considerably straighter and narrower than the blue
(Figure~\ref{gmrM}).  However, it is not a simple power law:  it is 
shallower between $-20.5>M_r>-22.5$ than at either the fainter or 
brighter ends (Figure~\ref{gmrEtypes} and Table~\ref{gmrMtable}).  
This curvature is not due to contamination by later morphological 
types at the faint end (Figure~\ref{gmrMorph}).  It also does not 
depend on whether one uses {\tt Petrosian} or {\tt model} colors 
(Figure~\ref{gmrMPM}); although color-gradients mean that the scale 
on which the color is defined does lead to small quantitative 
differences.  The curvature is also robust to (reasonable changes in) 
the choice of $k$-correction provided one properly accounts for 
evolution (Figure~\ref{gmrEvolM}).  Unless care is taken to account for 
it, this curvature may be confused with evolution in magnitude limited 
surveys (discussion following Figure~\ref{gmrEvolM}).  
All these properties of the color-magnitude relation are also true 
of the color-stellar mass relation (Figures~\ref{gmrEtypes}, \ref{gmrM}, 
\ref{gmrMs} and Tables~\ref{gmrMtable} and~\ref{gmrMsredblue}), and 
the color-$R_e$ relation (Figure~\ref{gmrSize}).  

The curvature towards redder colors at the brightest ($M_r\le -22.5$), 
most massive ($\log_{10}(M_*/M_\odot)\ge 11.3$) end is evident at fixed 
age and metallicity, suggesting that it is not driven by stellar 
population effects (Figure~\ref{fig|AgeMet}).  
In contrast, the color-$\sigma$ relation shows no curvature at 
high $\sigma > 150 \, {\rm km~s}^{-1}$ (Figure~\ref{gmrSigma}).  
The fact that there is no feature at the largest $\sigma$, despite 
clear features in the scalings with $M_*$, has strong implications 
for models of the assembly histories of massive galaxies.

Skelton et al. (2009) have argued that the change from a steeper 
slope at low luminosities to a shallower one at $M_r<-20.5$ is due 
to a change in formation histories.  They associate the shallower slope 
with recent major dry mergers which are expected to increase the 
luminosity and stellar mass without changing the color significantly.  
Since such mergers are expected to leave the velocity dispersion 
unchanged, that fact that there is no curvature in the 
color-$\sigma$ relation (Figure~\ref{gmrSigma}) seems in striking 
agreement with the dry major merger hypothesis.  In addition, dry 
major mergers are expected to increase the size in proportion to 
the mass, and we do see some flattening in the color-$R_e$ relation 
(Figure~\ref{gmrSize}).  
However, if the flattening at intermediate luminosities (and stellar 
masses, and sizes), with no curvature in the color-$\sigma$ relation is 
indeed due to major dry mergers, then it seems difficult for such a 
scenario to explain the steepening at even higher luminosities 
($M_r<-22.5$ or $\log (M_*/M_\odot) > 11.3$), even though these are 
precisely the objects for which the dry merger hypothesis is most 
commonly invoked.  

Therefore, we discussed three models that are compatible with our 
measurements:  one in which major mergers dominate the mass growth 
at $M_* > 2\times 10^{11}M_\odot$ (Figure~\ref{fig|Mssigma1}), 
another in which mergers are both major and minor, but the minor 
mergers at these largest masses contribute to the intracluster light 
(Figure~\ref{fig|Mssigma2}), and a third in which the reddest most 
massive objects today, which happen to also be the oldest, formed 
from major and minor mergers of the oldest, reddest objects in the 
past (Figure~\ref{fig|Mssigma3}), whereas the bluest objects formed 
from minor (but not major) mergers of blue objects.  Observations of 
the scatter and environmental dependence of the color-$M_*$ relation 
at $z\sim 1$, and of the color-$R_e$ relation at intermediate sizes, 
will discriminate between these models.  (The color-$\sigma$ relation 
is useful too; we are assuming it is harder to measure at high $z$.)

Such tests, e.g., using the thickness of the red sequence to constrain 
the formation histories of early-type galaxies, must be done with care.   
This is because although samples defined by cuts in concentration alone 
may provide a reliable estimate of the mean shape of the red sequence, 
they provide a bad estimate of the thickness (see Figures~\ref{gmrEtypes} 
and~\ref{gmrM}).  In particular, the red sequence in such samples is 
thicker at fainter luminosities, because of contamination by edge-on 
galaxies.  Appendix~\ref{2gs} shows that double-Gaussian fits to the 
bimodal color-magnitude relation, while purely statistical, provide 
a simple way of correcting approximately for this contamination.  
For example, at intermediate luminosities (i.e., around $L_*$), 
this procedure correctly assigns the reddest objects to the blue cloud, 
rather than to the red sequence (Figures~\ref{histgmr} and~\ref{histgmrMs}).  
Appendix~\ref{fukugita} shows that such objects tend to be edge-on 
spirals, and can be a significant source of contamination if one 
simply defines the red sequence by a straight color cut 
(Figure~\ref{gmrMorph}).  While they can also easily be removed by 
a cut on axis ratio (e.g. require $b/a \ge 0.6$), cutting on 
concentration index instead does not remove these objects 
(compare Figures~\ref{histgmrab} and~\ref{histgmrci}).  

In contrast to samples defined by color or concentration, the width of 
the red-sequence defined by double-Gaussian fits is independent of 
luminosity.  In our dataset, we find this width to be $0.033$~mags 
(Table~\ref{gmrMredblue}).  
Since measurement errors are of order 0.02~mags, the intrinsic width may 
be more like $0.026$~mags.  Our results suggest that, to obtain results 
which are less likely to be biased by contamination, it is this width 
which should be compared with the analogous quantity in higher redshift 
samples.  
On the other hand, we found that the double-Gaussian decomposition 
was not able to account for about 10\% of the objects at luminosities 
below $M_r<-20.5$; these objects tended to populate the green valley 
between the red and blue sequences.  So, if the double-Gaussian fits 
are to be used at higher redshift, one must first check that such 
objects are not much more common than they are at $z\sim 0.1$.  

Our models assume that massive objects have experienced major mergers 
since $z\sim 1$, meaning that the total stellar mass in early-types 
with $M_*>2\times 10^{11}M_\odot$ today must have been smaller by about 
a factor of 2 at $z\sim 1$.  It is not clear that this is consistent 
with current constraints.  E.g., although Faber et al. (2007) claim 
that the number density of early-types has increased by a factor of 
at least two since $z\sim 1$, and Matsuoka \& Kawara (2010) argue 
that the number density of objects with $M_*>2\times 10^{11}M_\odot$ 
has increased by an order of magnitude since $z\sim 1$, 
Brown et al. (2007), Wake et al. (2008) and Cool et al. (2008) claim 
that the mass growth since $z\sim 0.7$, for objects with 
$M_*>10^{11}M_\odot$, must have been less than 50\%.  
Eliche-Moral et al. (2010) argue that some of the discrepancy between 
these two claims is due to the difference between how the samples were 
defined.  However, most of these constraints are based on 
parametrizations of the $z\sim 0$ stellar mass function which may 
have underestimated the true abundance at $M_*>2\times 10^{11}M_\odot$ 
by 50\% (see Bernardi et al. 2010a and references therein).  If the 
true local abundance is indeed larger, then major mergers may be 
{\em required} to reconcile the $z\sim 1$ counts with those at $z\sim 0$.  

Finally, it is interesting to ask how BCGs, which are amongst the most 
massive objects in the local universe, fit into this picture?  
Although we do not show them explicitly here, they define a similar 
color-$M_*$ relation (and other relations as those shown in Figure~1 
of Bernardi et al. 2010b) for $\log (M_*/M_\odot) > 11.3$.  
However, there are some important differences:  
compared to non-BCGs of similar mass or luminosity, their colors are 
slightly redder (Figure~\ref{fig|Mssigma3}, and Roche et al. 2010), 
they have smaller color gradients (Roche et al. 2010), 
and slightly larger sizes (Bernardi 2009).  
Whereas the first two suggest merger histories dominated by major 
mergers, consistent with their large masses, the fact that their 
sizes are larger suggests more size growth than is usually associated 
with major mergers.  This suggests that although major mergers erased 
their color gradients at some higher redshift, minor mergers have puffed 
up their sizes, decreased their velocity dispersions further and 
contributed to the formation of ICL at lower redshift (Bernardi 2009).



\section*{Acknowledgments}
We are grateful to Simona Mei for a very helpful reading of our 
manuscript.  
MB thanks Meudon Observatory, and RKS thanks the IPhT at CEA-Saclay, 
for their hospitality during the course of this work.  
MB is grateful for support provided by NASA grant ADP/NNX09AD02G; 
FS acknowledges support from the Alexander von Humboldt Foundation;  
RKS is supported in part by NSF-AST 0908241.

Funding for the Sloan Digital Sky Survey (SDSS) and SDSS-II Archive has been
provided by the Alfred P. Sloan Foundation, the Participating Institutions, the
National Science Foundation, the U.S. Department of Energy, the National
Aeronautics and Space Administration, the Japanese Monbukagakusho, and the Max
Planck Society, and the Higher Education Funding Council for England. The
SDSS Web site is http://www.sdss.org/.

The SDSS is managed by the Astrophysical Research Consortium (ARC) for the
Participating Institutions. The Participating Institutions are the American
Museum of Natural History, Astrophysical Institute Potsdam, University of Basel,
University of Cambridge, Case Western Reserve University, The University of
Chicago, Drexel University, Fermilab, the Institute for Advanced Study, the
Japan Participation Group, The Johns Hopkins University, the Joint Institute
for Nuclear Astrophysics, the Kavli Institute for Particle Astrophysics and
Cosmology, the Korean Scientist Group, the Chinese Academy of Sciences (LAMOST),
Los Alamos National Laboratory, the Max-Planck-Institute for Astronomy (MPIA),
the Max-Planck-Institute for Astrophysics (MPA), New Mexico State University,
Ohio State University, University of Pittsburgh, University of Portsmouth,
Princeton University, the United States Naval Observatory, and the University
of Washington.

\appendix

\section{Robustness to changes in how the red-sequence is defined}


\subsection{Double-gaussian fits to the bimodality}\label{2gs}

\begin{figure*}
 \centering
 \includegraphics[width=0.69\hsize]{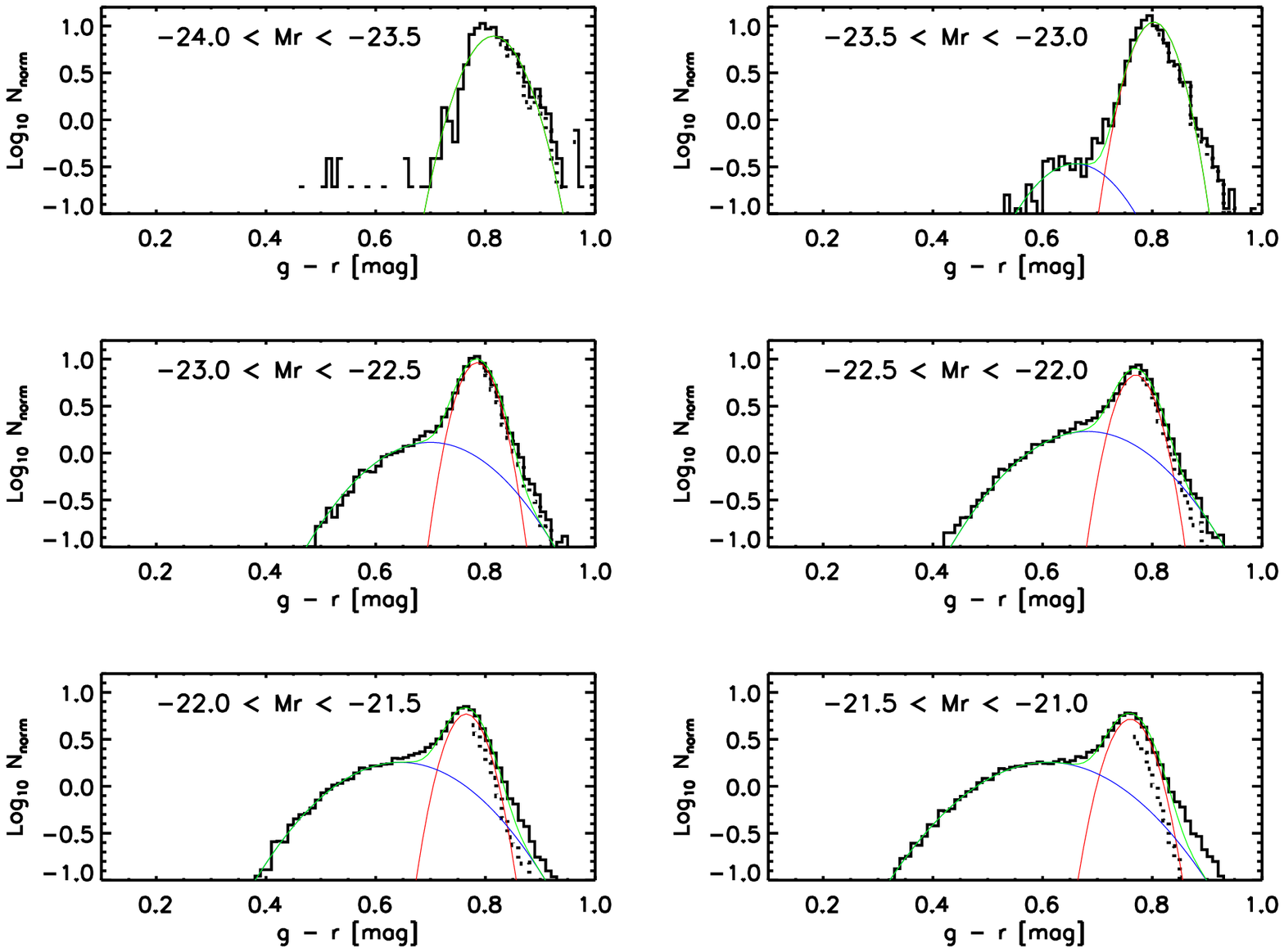}
 \includegraphics[width=0.69\hsize]{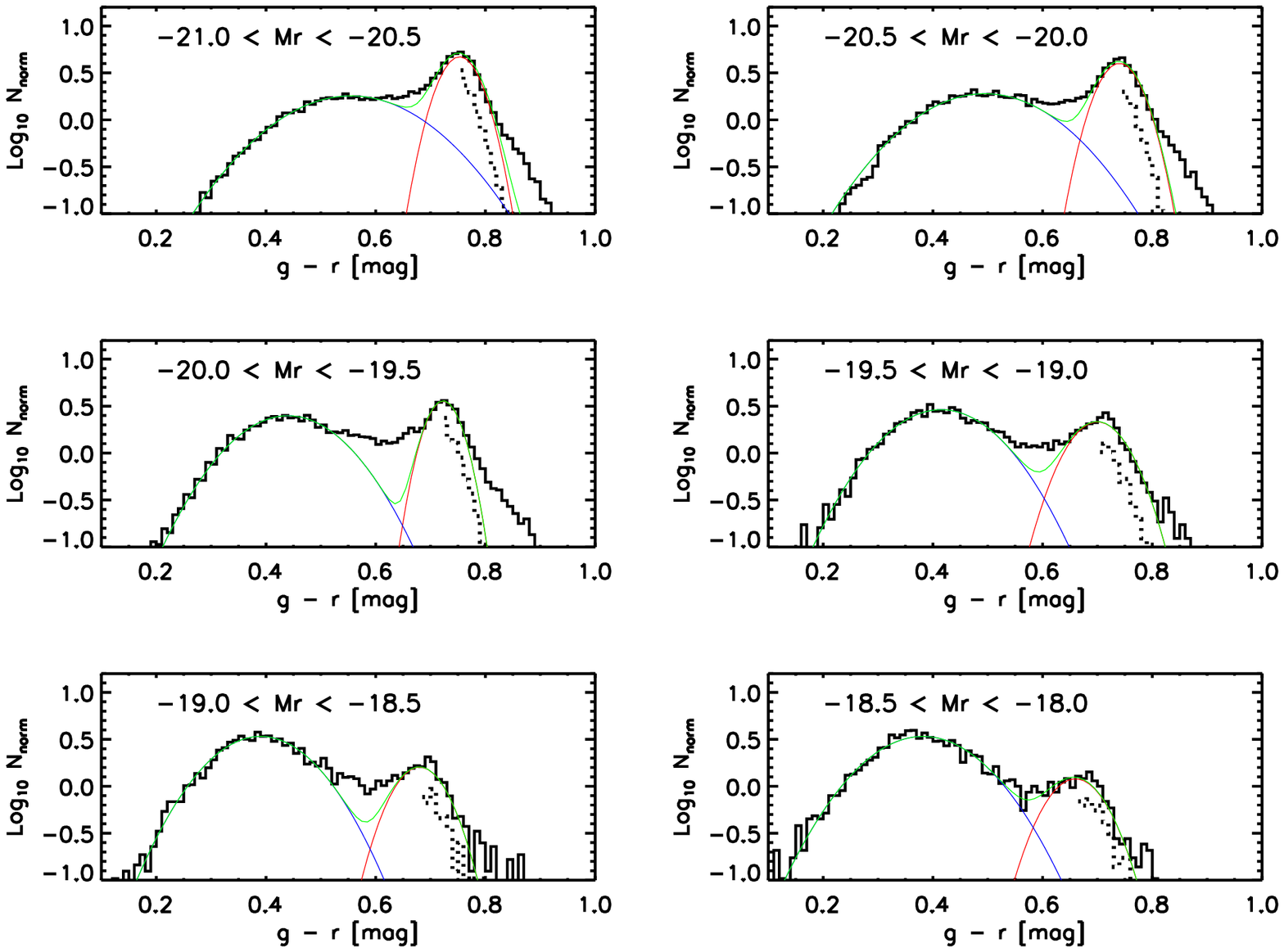}
 \caption{Double-gaussian fits to the bimodal color-magnitude relation 
          provide a good description except in the range
          $-18.5\le M_r\le -20.5$ (the ``green valley''). 
          Note that, at intermediate/high luminosities, the reddest 
          objects are actually associated with the red tail of the blue 
          component, consistent with the physical expectation that 
          the red sequence may be contaminated by edge-on disks at these 
          luminosities.  The dotted line shows the red-end distribution 
          of galaxies (i.e. objects redder than the mean of the red 
          Gaussian component) that also have $b/a > 0.6$.  Clearly, 
          the reddest objects have smaller $b/a$, consistent with their 
          being edge-on disks. }
 \label{histgmr}
\end{figure*}

\begin{figure*}
 \centering
 \includegraphics[width=0.79\hsize]{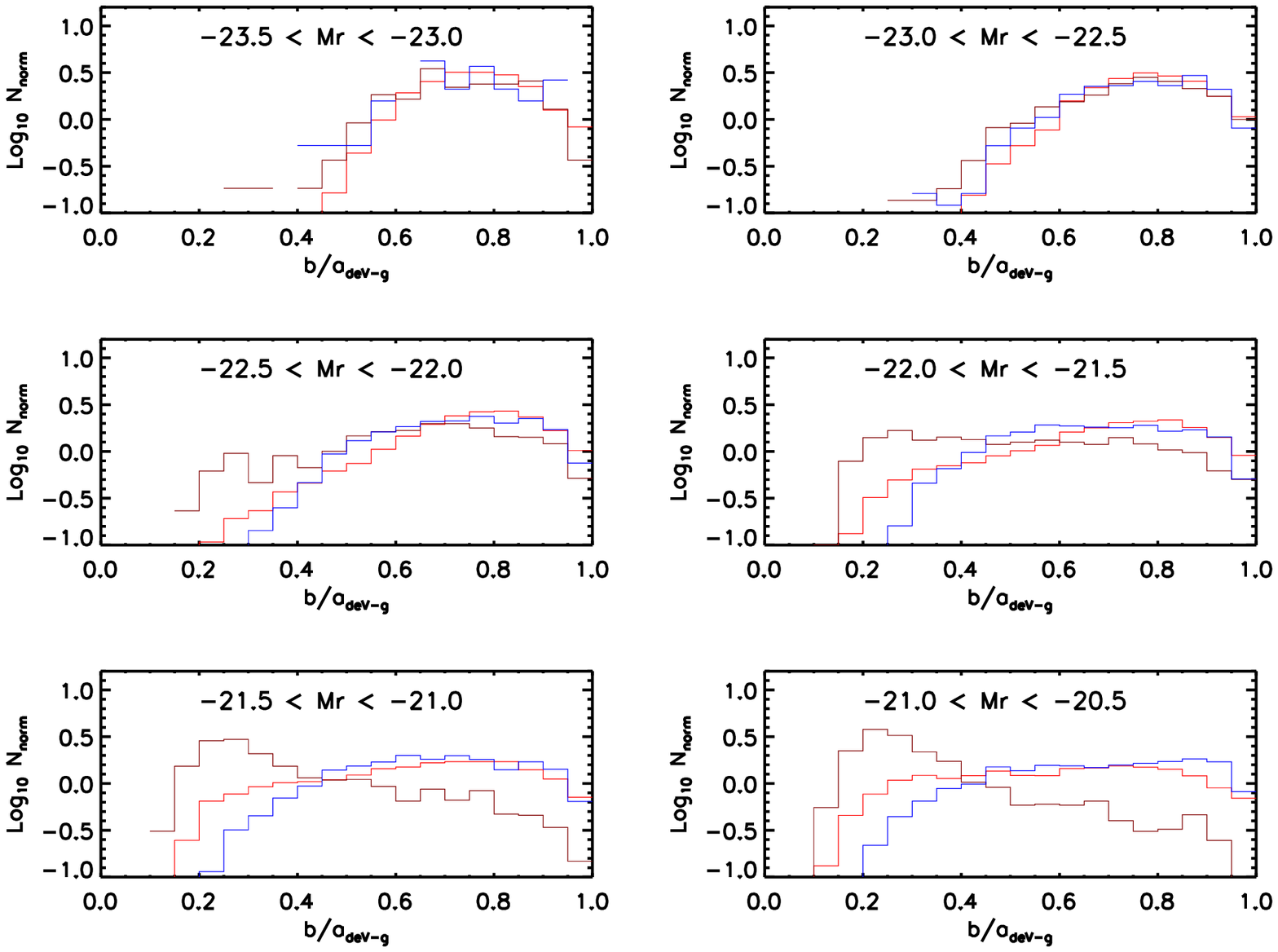}
 \includegraphics[width=0.79\hsize]{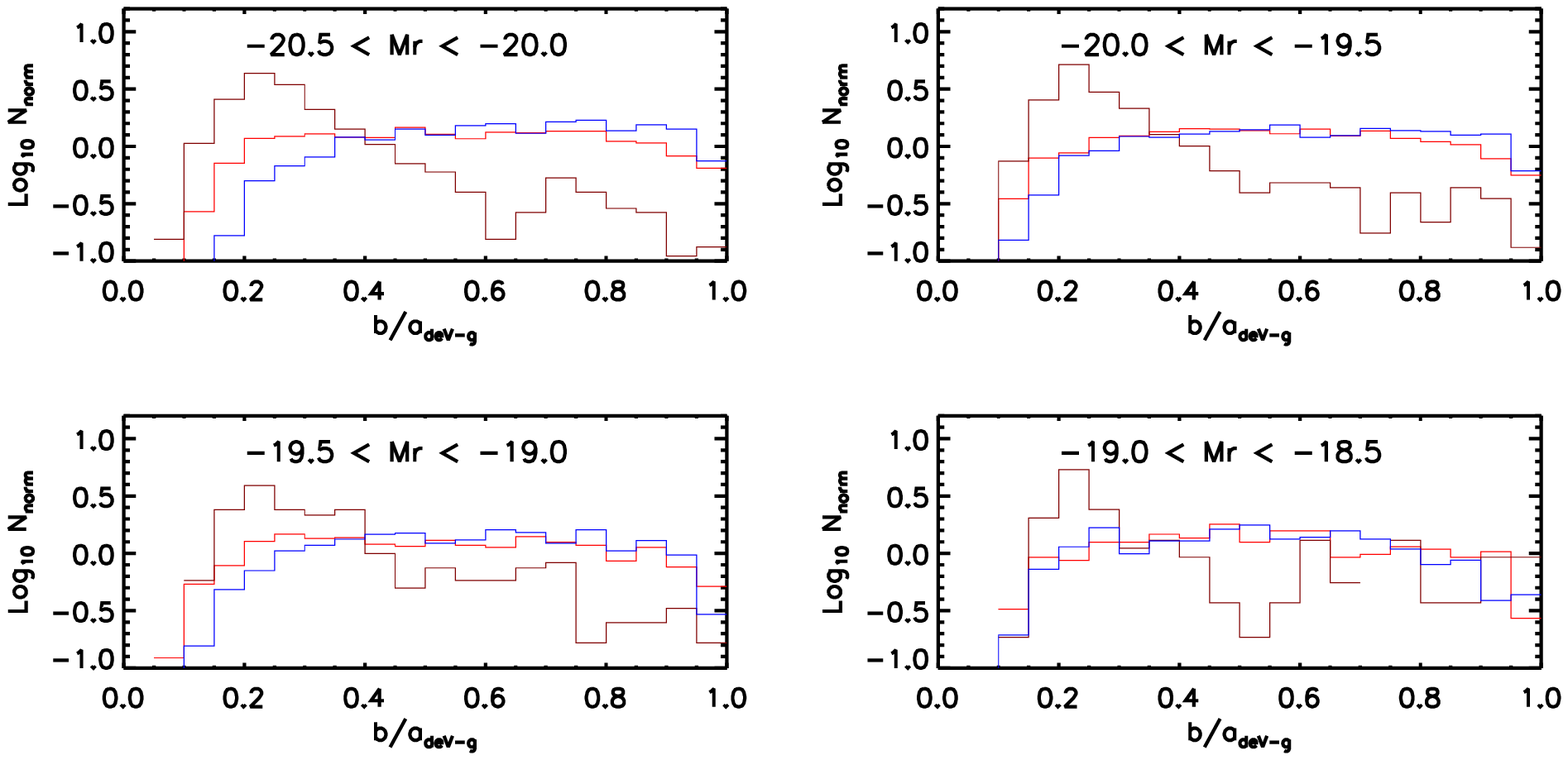}
 \vspace{-1.3in}
 \caption{Distribution of axis ratio $b/a$ for objects that are within 
         $0.025$~mags of the peak of the blue component (blue), 
         $0.025$~mags of the peak of the red sequence (red), 
         and 0.1~mags redder than the peak of the red sequence (brown).  
         Compared to the other two populations, this final population has 
         an excess of small $b/a$ values:  many objects which are 
         significantly redder than the mean red sequence tend to be 
         edge-on disks. }
 \label{histgmrab}
\end{figure*}

\begin{figure*}
 \centering
 \includegraphics[width=0.79\hsize]{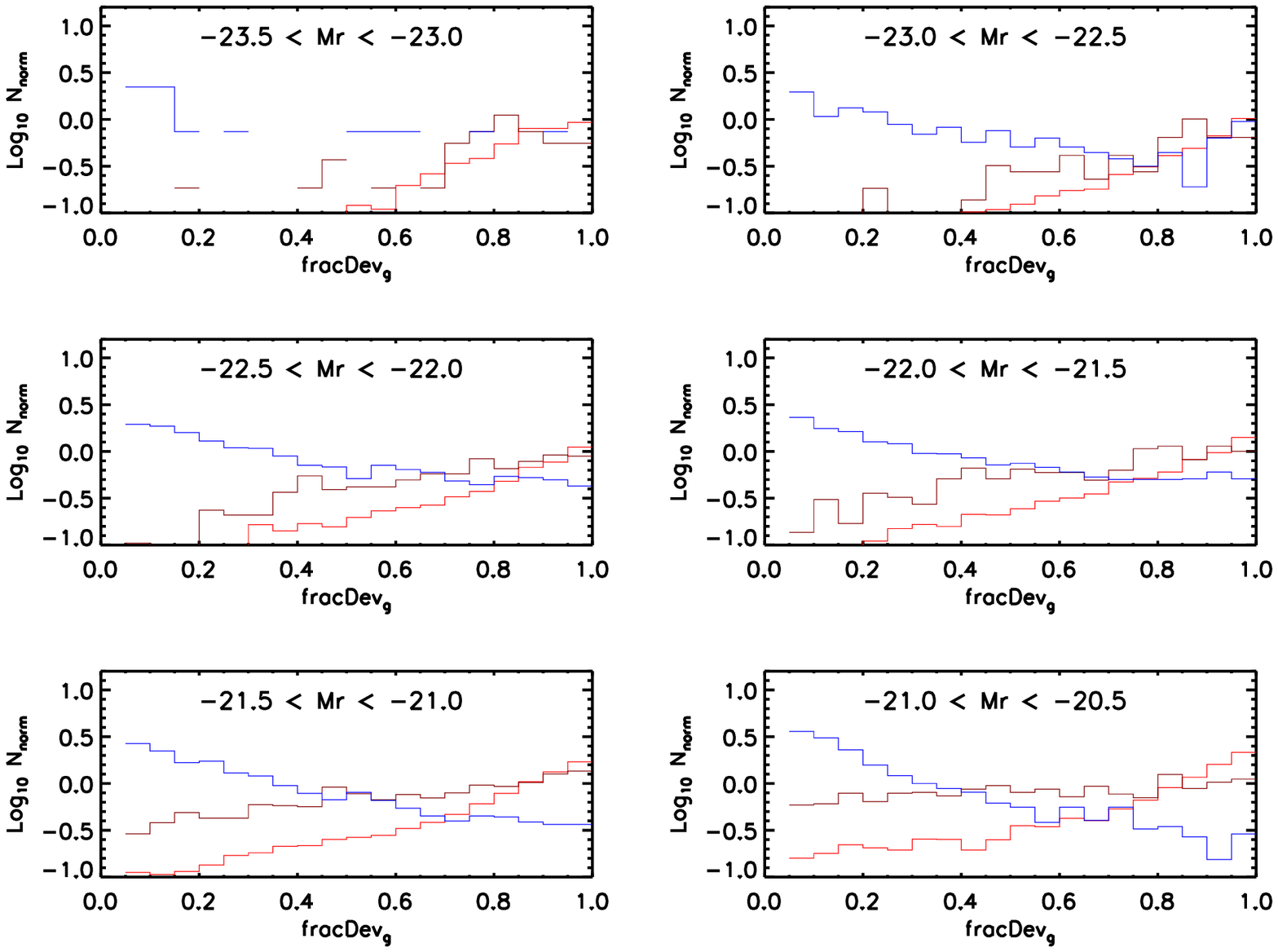}
 \includegraphics[width=0.79\hsize]{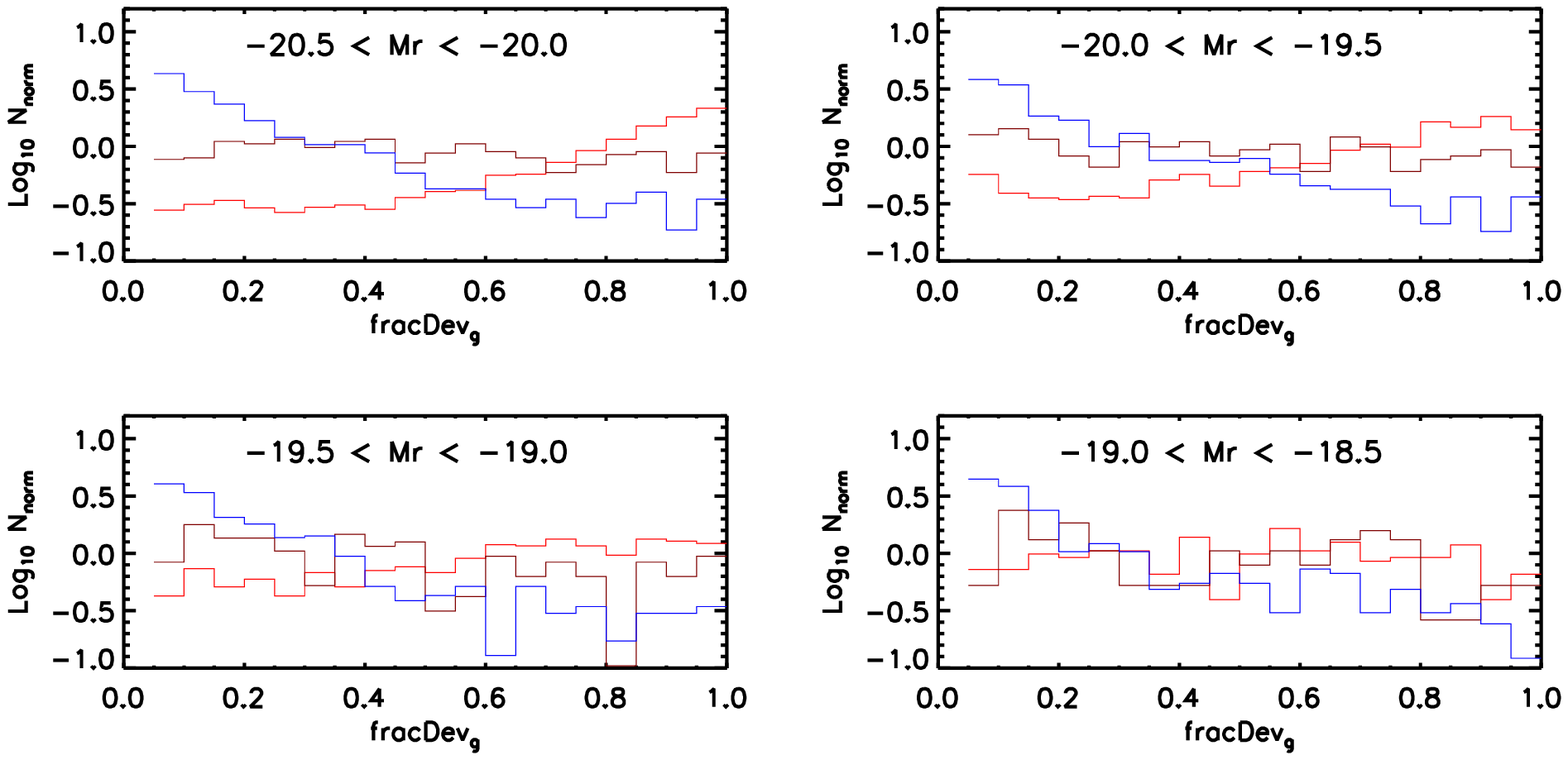}
 \vspace{-1.3in}
\caption{Distribution of {\tt fracDev}, i.e. the weight of the deVaucouleurs 
         component in the best composite model, for objects that are within 
         $0.025$~mags of the peak of the blue component (blue), 
         $0.025$~mags of the peak of the red sequence (red), 
         and 0.1~mags redder than the peak of the red sequence (brown).  
         Objects near the peak of the blue/red sequence tend to have 
         low/high values of {\tt fracDev}, as expected.
         However, objects which lie redward of the red sequence tend to 
         have small values of {\tt fracDev} more often than do objects 
         which lie at the peak of the red sequence.  }
 \label{histgmrfracdev}
\end{figure*}

\begin{figure*}
 \centering
 \includegraphics[width=0.79\hsize]{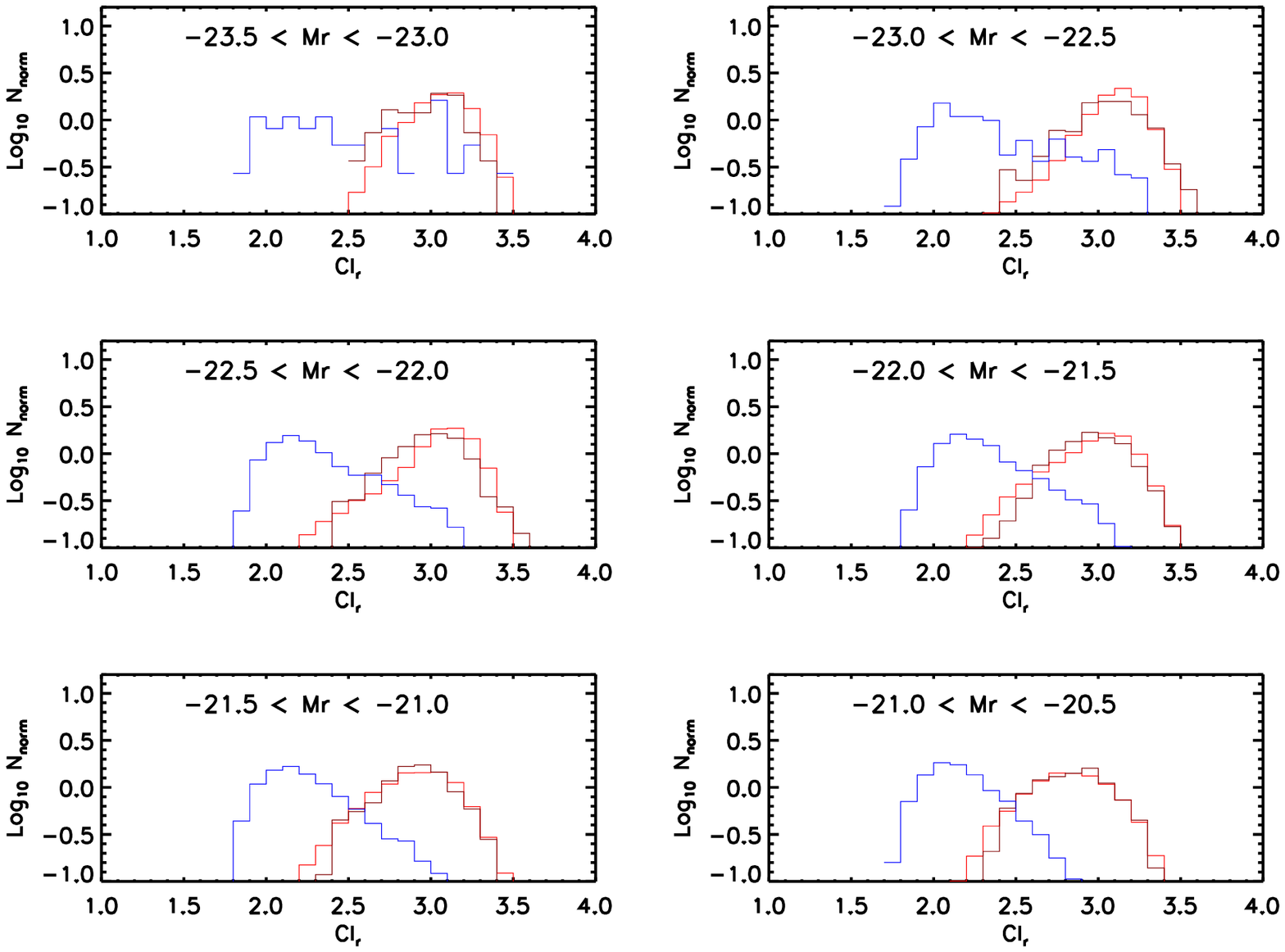}
 \includegraphics[width=0.79\hsize]{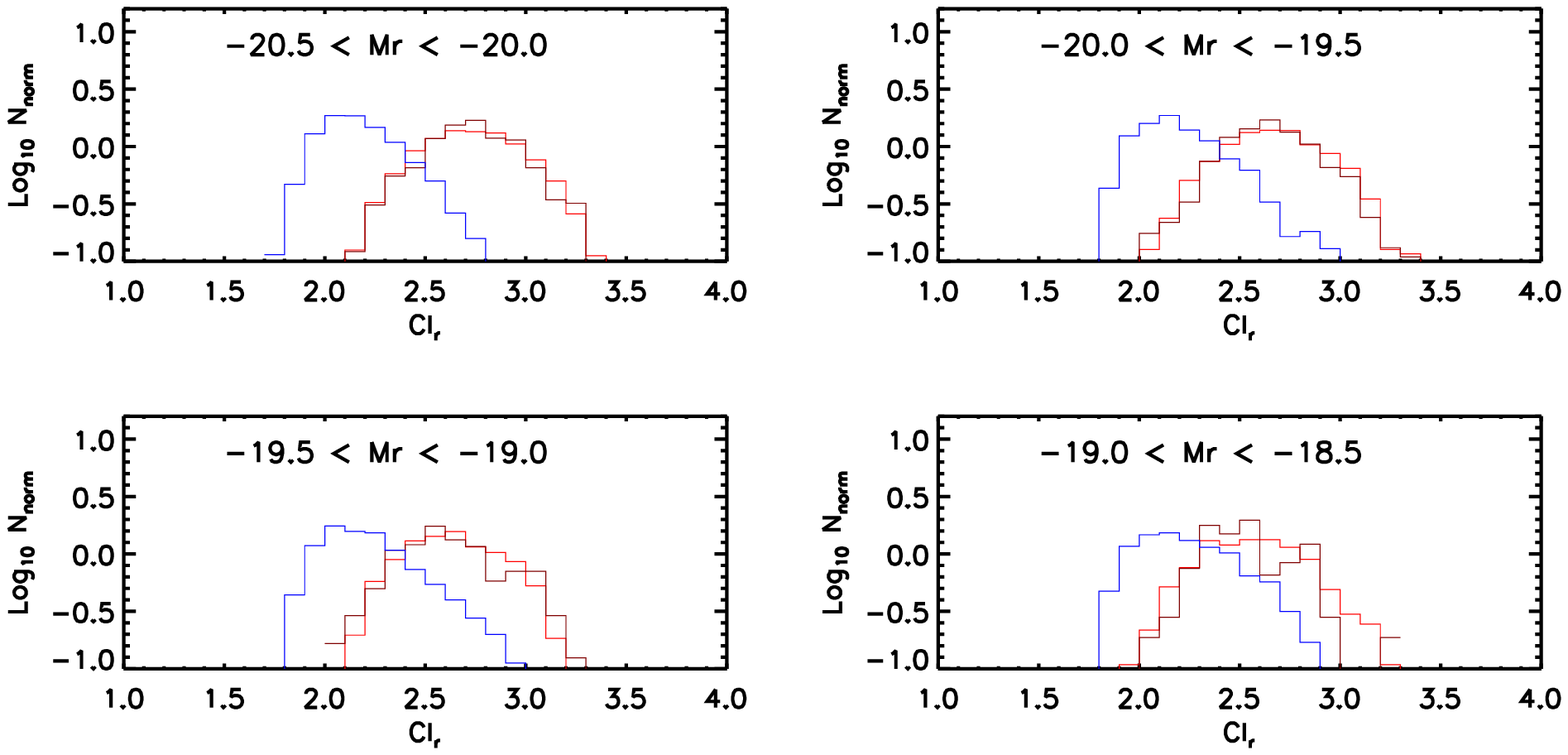}
 \vspace{-1.3in}
\caption{Distribution of the concentration index $C_r$ for objects 
         that are within 
         $0.025$~mags of the peak of the blue component (blue), 
         $0.025$~mags of the peak of the red sequence (red), 
         and 0.1~mags redder than the peak of the red sequence (brown).  
         The reddest objects, which the previous figures showed tend to 
         have small $b/a$ and {\tt fracDev}, have $C_r$ values which are 
         indistinguishable from those of genuine red sequence galaxies.  
         Hence, cuts on concentration are {\em not} a reliable way to 
         identify and eliminate such objects.  }
 \label{histgmrci}
\end{figure*}

\begin{figure*}
 \centering
 \includegraphics[width=0.69\hsize]{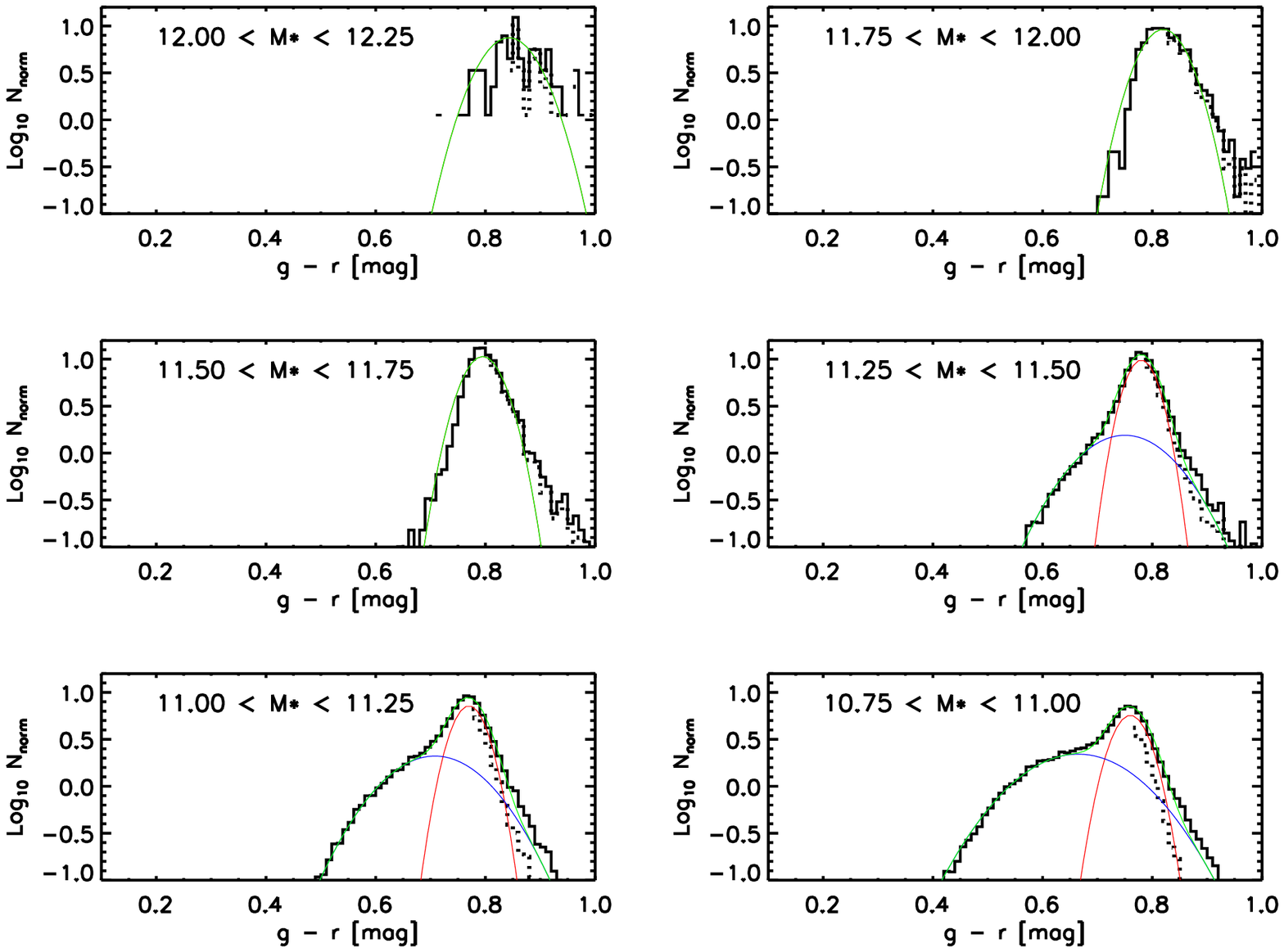}
 \includegraphics[width=0.69\hsize]{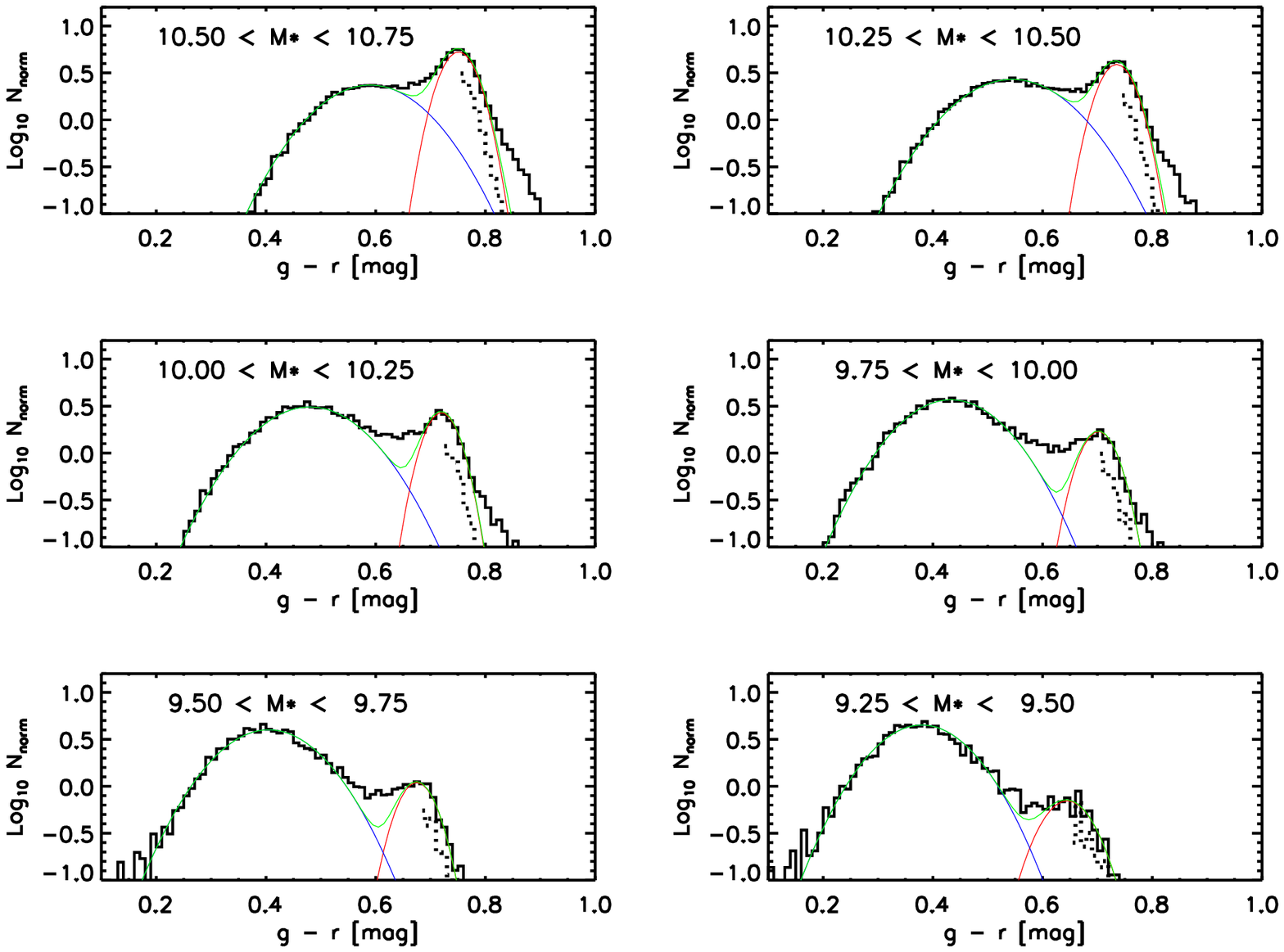}
 \caption{Same format as Figure~\ref{histgmr} but showing the double-gaussian fits to the bimodal color-$M_*$ relation.}
 \label{histgmrMs}
\end{figure*}


\begin{table*}
\caption[]{Parameters of double-Gaussian fits to the $g-r$ model color 
           distribution in narrow bins in $M_r$.}
\begin{tabular}{cccccccr}
 \hline
  $M_r$ & $g-r$ (RED) & $rms$ (RED) & $g-r$ (BLUE) & $rms$ (BLUE) & $\%$ (RED) &  $\%$ (BLUE) & N$_{\rm{gal}}$ \\
\hline
$-23.62$ & $ 0.815$ & $ 0.043$ & $ 0.000$ & $ 0.000$ & $ 1.000$ & $ 0.000$ & $  517$ \\
$-23.16$ & $ 0.803$ & $ 0.033$ & $ 0.660$ & $ 0.070$ & $ 0.910$ & $ 0.060$ & $ 4413$ \\
$-22.69$ & $ 0.785$ & $ 0.030$ & $ 0.700$ & $ 0.100$ & $ 0.692$ & $ 0.326$ & $16906$ \\
$-22.22$ & $ 0.770$ & $ 0.031$ & $ 0.682$ & $ 0.105$ & $ 0.528$ & $ 0.447$ & $38080$ \\
$-21.74$ & $ 0.765$ & $ 0.032$ & $ 0.645$ & $ 0.110$ & $ 0.473$ & $ 0.496$ & $53391$ \\
$-21.26$ & $ 0.760$ & $ 0.034$ & $ 0.610$ & $ 0.120$ & $ 0.443$ & $ 0.541$ & $50603$ \\
$-20.77$ & $ 0.753$ & $ 0.035$ & $ 0.555$ & $ 0.120$ & $ 0.412$ & $ 0.541$ & $39172$ \\
$-20.28$ & $ 0.740$ & $ 0.037$ & $ 0.495$ & $ 0.115$ & $ 0.371$ & $ 0.548$ & $25833$ \\
$-19.79$ & $ 0.723$ & $ 0.030$ & $ 0.440$ & $ 0.090$ & $ 0.271$ & $ 0.564$ & $14081$ \\
$-19.28$ & $ 0.700$ & $ 0.050$ & $ 0.415$ & $ 0.090$ & $ 0.269$ & $ 0.654$ & $ 8114$ \\
$-18.78$ & $ 0.680$ & $ 0.045$ & $ 0.390$ & $ 0.085$ & $ 0.180$ & $ 0.724$ & $ 4817$ \\
$-18.29$ & $ 0.660$ & $ 0.050$ & $ 0.382$ & $ 0.095$ & $ 0.150$ & $ 0.810$ & $ 2892$ \\
$-17.80$ & $ 0.630$ & $ 0.050$ & $ 0.365$ & $ 0.090$ & $ 0.150$ & $ 0.857$ & $ 1506$ \\
\hline 
\end{tabular}
\label{gmrMredblue} 
\end{table*}

\begin{table*}
\caption[]{Parameters of double-Gaussian fits to the $g-r$ model color 
           distribution in narrow bins in $\log_{10}M_\star$.}
\begin{tabular}{cccccccr}
 \hline
  Log$_{10}M_*/M_\odot$ & $g-r$ (RED) & $rms$ (RED) & $g-r$ (BLUE) & $rms$ (BLUE) & $\%$ (RED) & $\%$ (BLUE) & N$_{\rm{gal}}$ \\
\hline
$ 12.04$ & $ 0.843$ & $ 0.048$ & $ 0.000$ & $ 0.000$ & $ 1.000$ & $ 0.000$ & $   89$ \\
$ 11.81$ & $ 0.820$ & $ 0.040$ & $ 0.000$ & $ 0.000$ & $ 0.922$ & $ 0.000$ & $ 1313$ \\
$ 11.58$ & $ 0.795$ & $ 0.035$ & $ 0.000$ & $ 0.000$ & $ 0.939$ & $ 0.000$ & $ 7943$ \\
$ 11.35$ & $ 0.780$ & $ 0.028$ & $ 0.750$ & $ 0.080$ & $ 0.688$ & $ 0.311$ & $24943$ \\
$ 11.11$ & $ 0.770$ & $ 0.030$ & $ 0.708$ & $ 0.085$ & $ 0.541$ & $ 0.447$ & $48026$ \\
$ 10.88$ & $ 0.760$ & $ 0.032$ & $ 0.665$ & $ 0.100$ & $ 0.457$ & $ 0.551$ & $55556$ \\
$ 10.63$ & $ 0.751$ & $ 0.032$ & $ 0.590$ & $ 0.090$ & $ 0.425$ & $ 0.530$ & $45702$ \\
$ 10.39$ & $ 0.735$ & $ 0.032$ & $ 0.545$ & $ 0.095$ & $ 0.313$ & $ 0.643$ & $31315$ \\
$ 10.14$ & $ 0.720$ & $ 0.030$ & $ 0.480$ & $ 0.090$ & $ 0.203$ & $ 0.699$ & $19064$ \\
$  9.89$ & $ 0.702$ & $ 0.032$ & $ 0.433$ & $ 0.085$ & $ 0.136$ & $ 0.788$ & $11693$ \\
$  9.64$ & $ 0.675$ & $ 0.033$ & $ 0.405$ & $ 0.085$ & $ 0.091$ & $ 0.852$ & $ 7077$ \\
$  9.39$ & $ 0.645$ & $ 0.045$ & $ 0.380$ & $ 0.080$ & $ 0.079$ & $ 0.902$ & $ 4388$ \\
$  9.15$ & $ 0.610$ & $ 0.045$ & $ 0.355$ & $ 0.080$ & $ 0.062$ & $ 0.882$ & $ 2463$ \\
\hline 
\end{tabular}
\label{gmrMsredblue} 
\end{table*}

The distribution of colors at fixed $L_r$ is well-known to be bimodal.  
The smooth curves in Figure~\ref{histgmr} show the result of fitting 
the sum of two gaussian components to the $g-r$ distribution at each $L_r$ 
(e.g. Baldry et al. 2004; Skibba \& Sheth 2009).
Note that the red sequence is considerably narrower than the bluer 
component.  The parameters of these fits are provided in 
Table~\ref{gmrMredblue}, and are used in the main text.  

Figure~\ref{histgmr} shows that, except in the range 
$-18.5\le M_r\le -20.5$ or $9.5\le \log M_*/M_\odot\le 10.25$ 
the double-gaussian is a good description of the measurements.  
However, at intermediate $L$ and $M_*$, it is unable to describe 
the transition region between the two populations.  
(Table~\ref{gmrMredblue} shows that, in this regime, 
the double-Gaussian decomposition only accounts for 90\% of the objects.)
Since this is fainter than the scales on which we see curvature 
in the color-magnitude relation, this is not a major concern.  
However, at slightly larger luminosities, the fits assign the reddest 
objects to the red tail of the blue component.  Is this a limitation 
of the statistical decomposition, or does it reflect something 
physical?  If it is physical, then this cautions against using 
sharp cuts in color to isolate early-type galaxies.  

The dotted line in Figure~\ref{histgmr} shows the red-end distribution 
of galaxies (i.e. objects redder than the mean of the red Gaussian 
component) with $b/a > 0.6$. This distribution is better fit by the 
red Gaussian component than by the red tail of the blue component.  
This shows that the objects which populate the extremely red tail of 
the blue Gaussian component tend to have small $b/a$.  

To address this further, Figures~\ref{histgmrab}--\ref{histgmrci} show 
the distribution of axis ratios $b/a$, and two measures of the shape 
of the light profile, {\tt fracDev} and concentration index, for 
objects that lie close to the peak of the blue and red sequences, 
and that lie 0.1~mags redward of the red sequence.  Notice that 
these reddest objects tend to have small values of $b/a$.  This 
suggests that they are edge-on disks, something which is corroborated 
by the fact that the distribution of {\tt fracDev} also extends to 
smaller values, characteristic of late-type galaxies, than it does 
for objects on the red sequence.  The distribution of concentrations, 
however, is just like that for objects on the red sequence, but note 
that there is significant overlap in between the distributions defined 
by the red and blue sequences.  

That the reddest objects at intermediate luminosity are late-type 
galaxies is also seen in Figure~13 of Bernardi et al. (2010a) which 
shows how the bimodal color-magnitude distribution is built up by 
objects of different morphological type.  Clearly, the reddest 
objects at $M_r\ge -22.5$ are primarily of type Sa and later -- they 
are not ellipticals.  In particular, they are not what one typically 
associates with the red sequence.  
That edge-on disks are amongst the reddest objects is not surprising.  
However, given the wide-spread use of concentration as a way of 
identifying red-sequence galaxies, our finding that concentration 
does such a poor job of identifying edge-on disks is disturbing.  
Our results caution against using sharp cuts in color or 
concentration for identifying early-type galaxies.  

\begin{figure*}
 \centering
 \includegraphics[width=0.375\hsize]{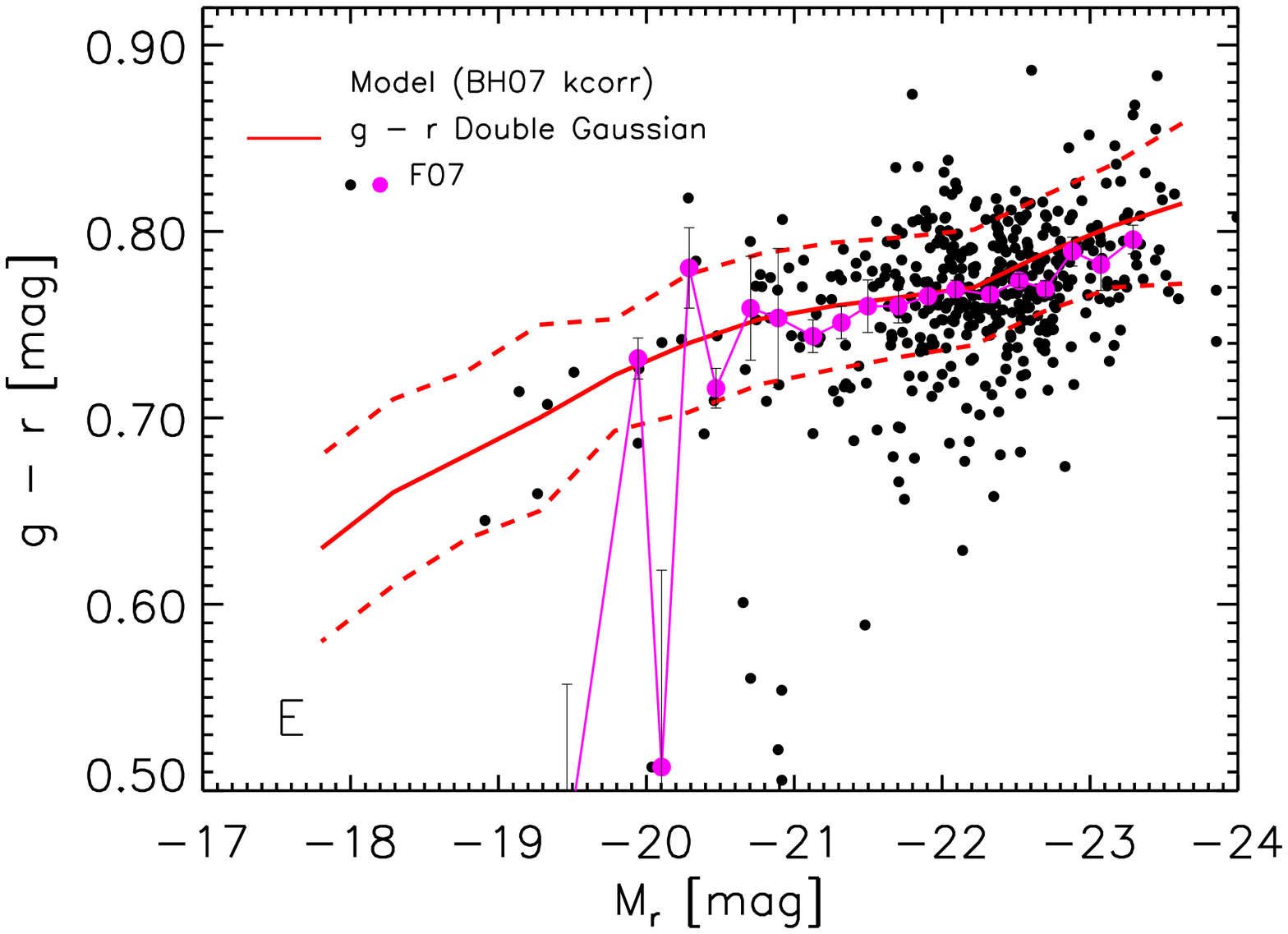}
 \includegraphics[width=0.375\hsize]{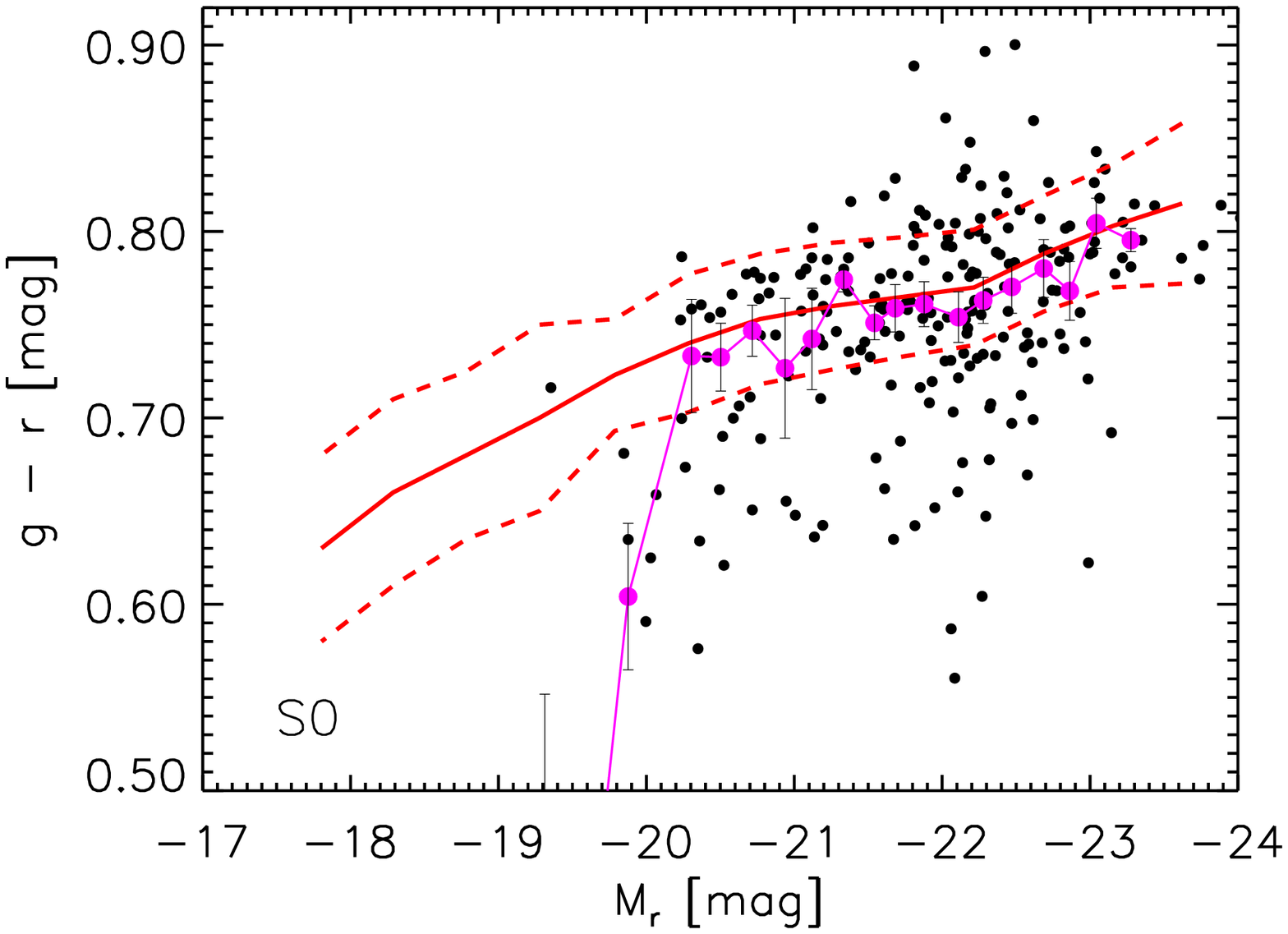} 
 \includegraphics[width=0.375\hsize]{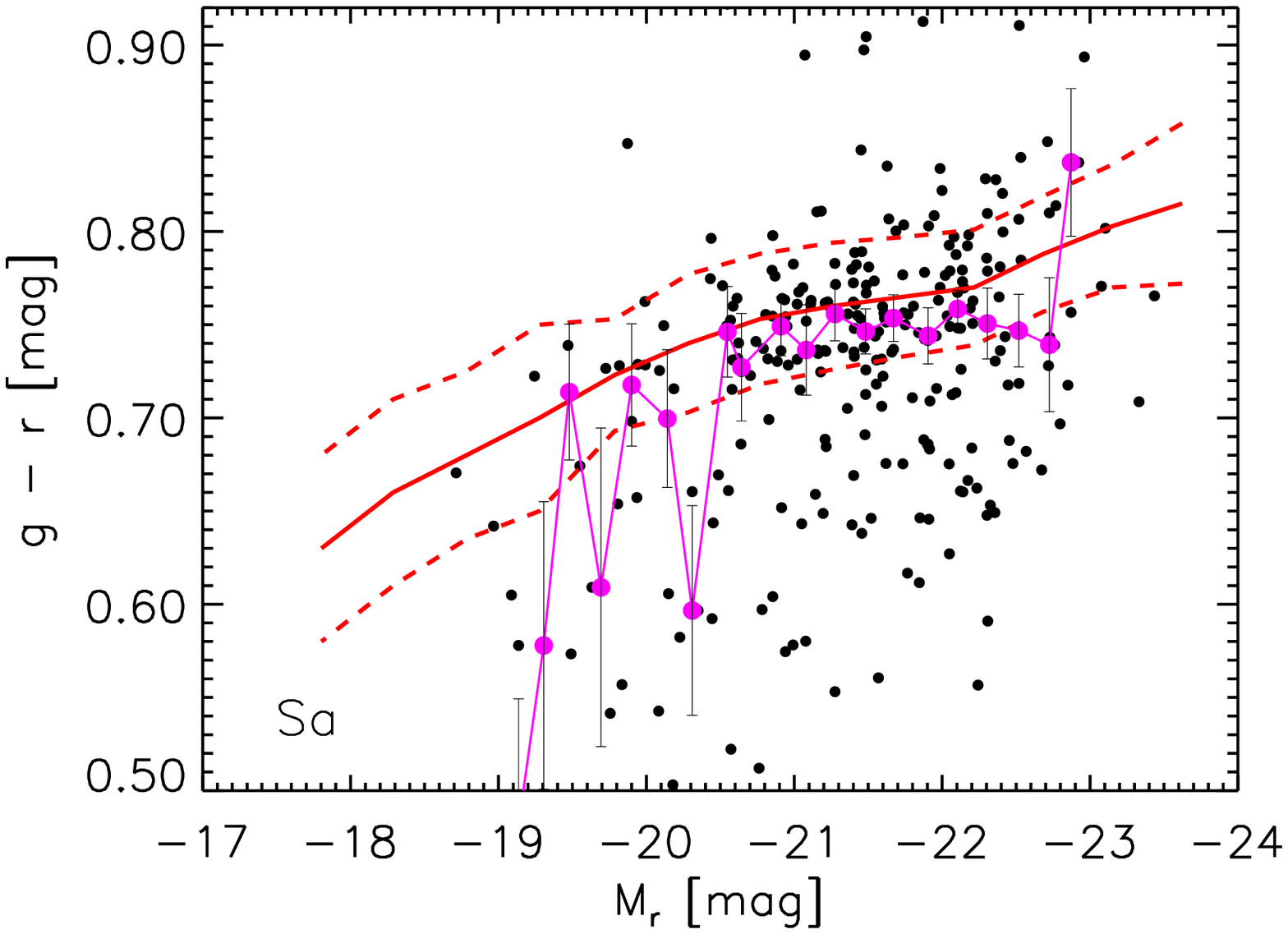}
 \includegraphics[width=0.375\hsize]{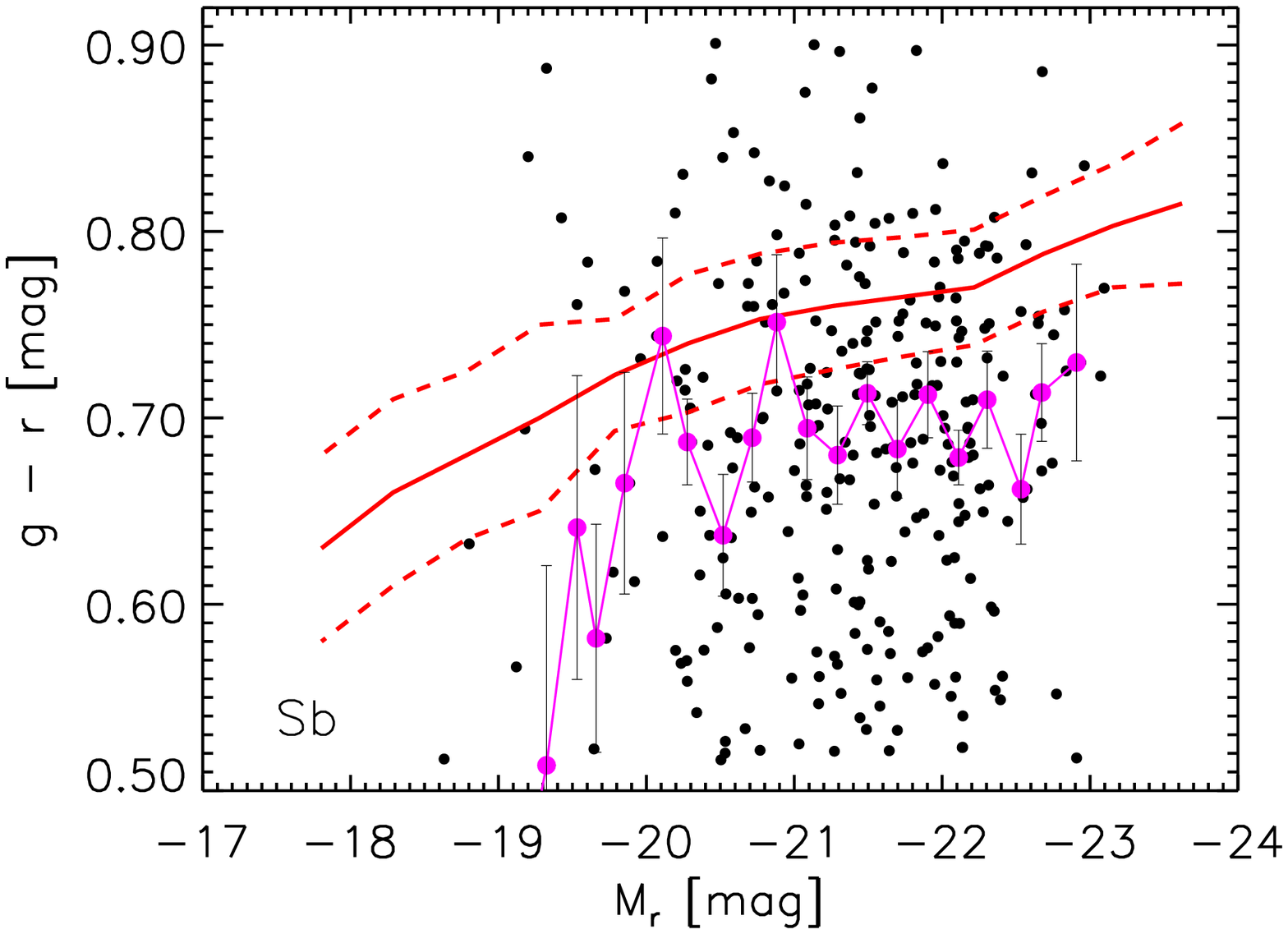}  
 \caption{Color-magnitude relation in the Fukugita et al. (2007) sample
          for types E, S0, Sa, and Sb.  Small filled circles show the 
          objects, large filled circles connected by jagged line, show 
          the mean color in bins of $M_r$ having width 0.25~mags.  
          Thick solid and associated dashed lines (same in all panels), 
          show the red sequence defined by our double-Gaussian fits to 
          the full SDSS sample (see Table~\ref{gmrMredblue}).  
          Sas and Sbs dominate the numbers redward of the red sequence.}
 \label{gmrMorph}
\end{figure*}

\begin{figure*}
 \centering
 \includegraphics[width=0.375\hsize]{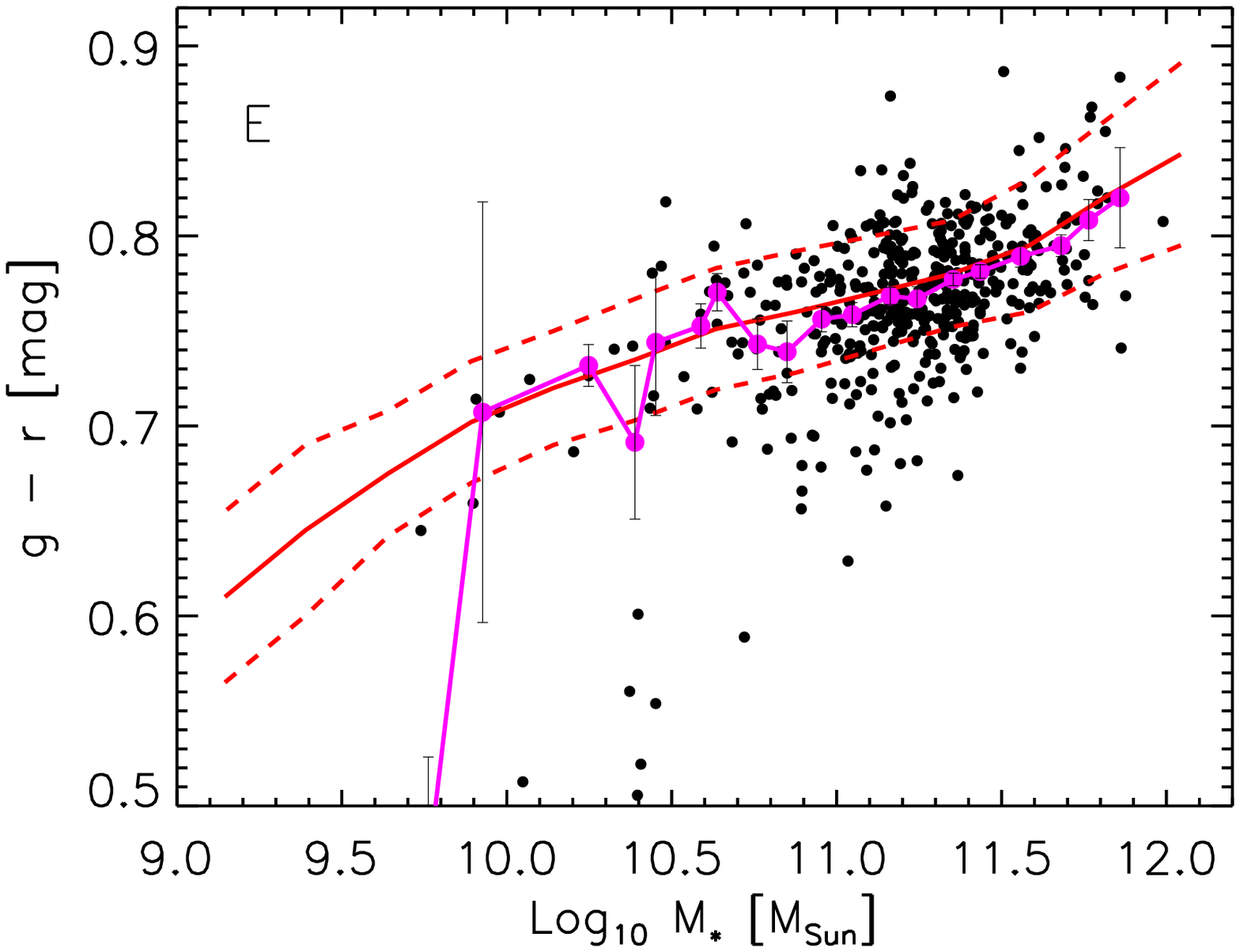}
 \includegraphics[width=0.375\hsize]{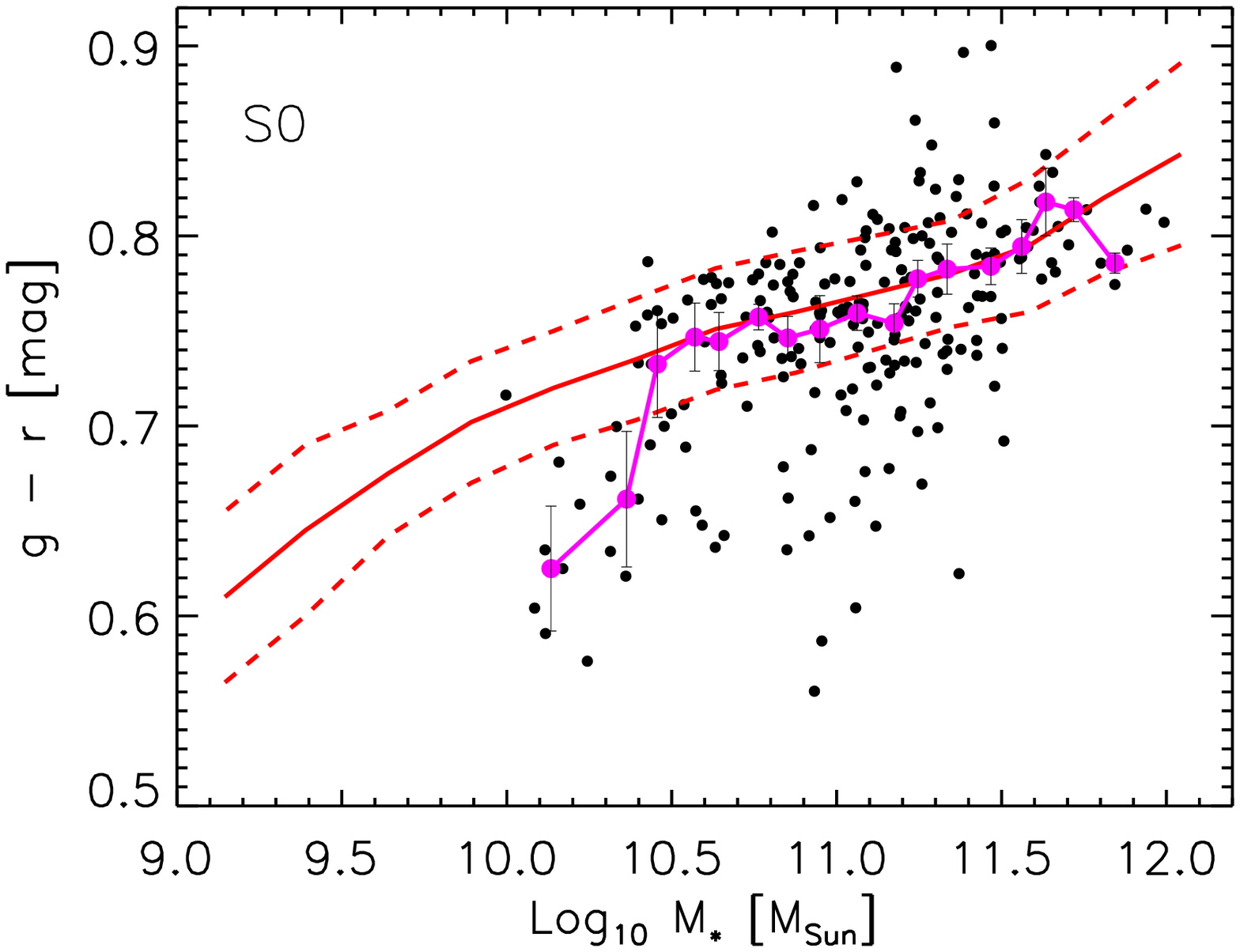} 
 \includegraphics[width=0.375\hsize]{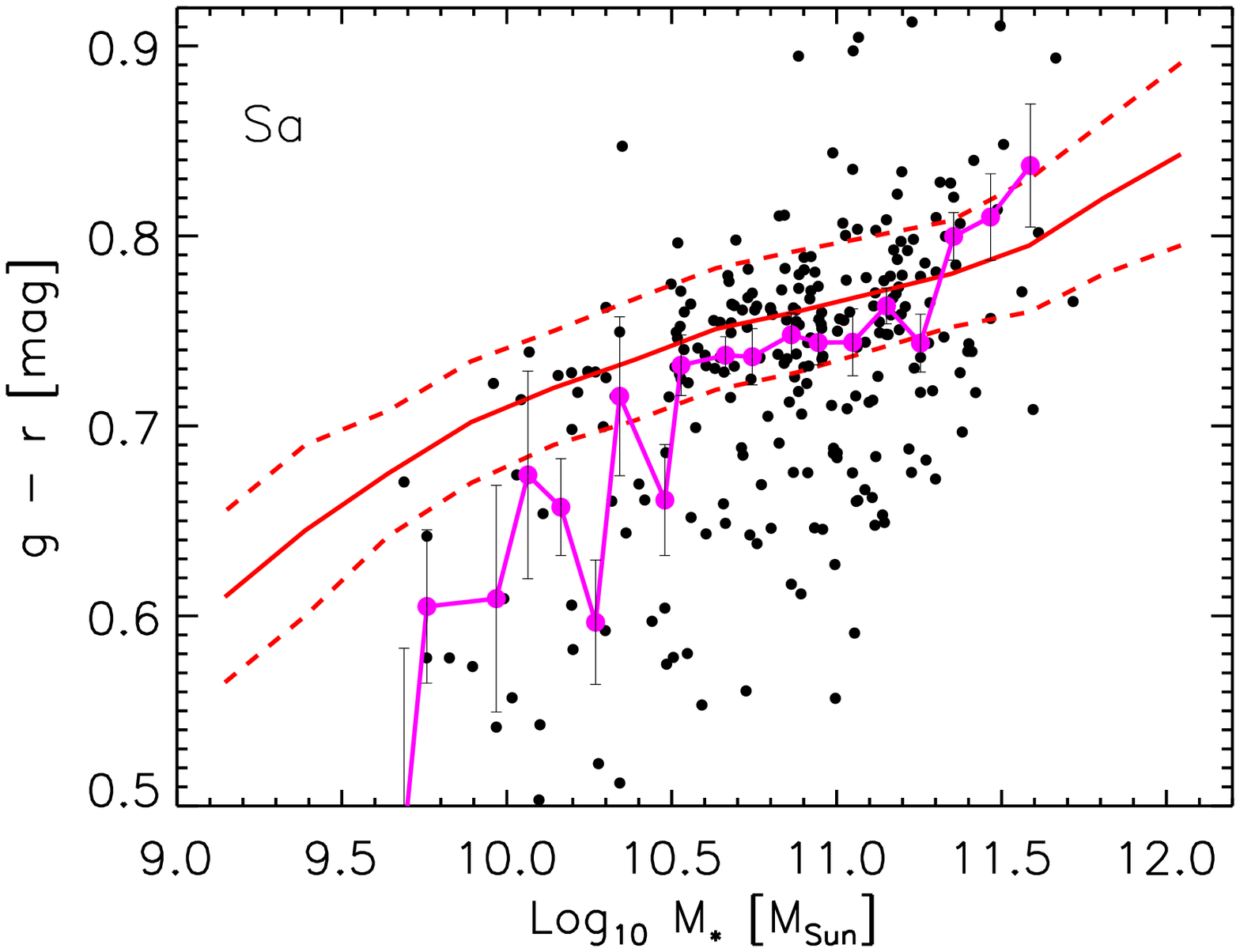}
 \includegraphics[width=0.375\hsize]{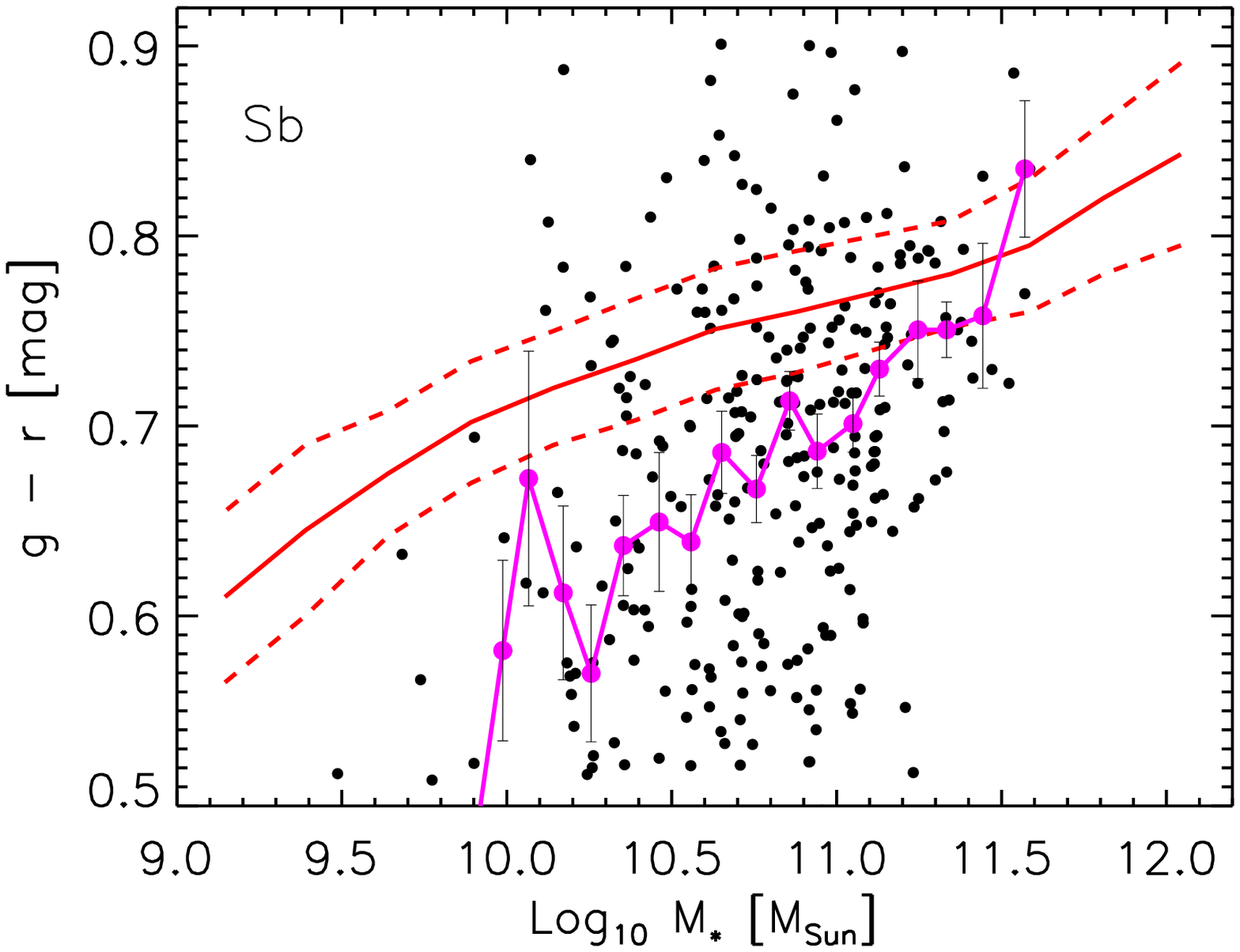}  
 \caption{Same as Figure~\ref{gmrMorph}, but with luminosity 
          replaced by stellar mass. 
          Table~\ref{gmrMsredblue} provides the parameters of the 
          red sequence defined by double-Gaussian fits, shown here 
          as the solid red line.}
 \label{gmrMs}
\end{figure*}

\subsection{Dependence on morphology}\label{fukugita}
One of our goals is to compare measurements on subsamples defined 
by relatively simple criteria (e.g. concentration index, bimodality) 
with morphologically selected subsamples.  To this end, we use the 
morphological classification provided by Fukugita et al. (2007).  
Briefly, Fukugita et al. have provided morphological classifications 
(Hubble type T) for a subset of 2253 SDSS galaxies brighter than 
$m_{\tt Pet}=16$ in the $r-$band, selected from 230 deg$^2$ of sky. 
Of these, 1866 have spectroscopic information.
Here, we group galaxies classified
with half-integer T into the smaller adjoining integer bin 
(except for the E class; see also Huang \& Gu 2009 and 
Oohama et al. 2009).  
In the following, we set E (T = 0 and 0.5), S0 (T = 1), 
Sa (T = 1.5 and 2), Sb (T = 2.5 and 3), and 
Scd (T = 3.5, 4, 4.5, 5, and 5.5).
This gives a fractional morphological mix of 
(E, S0, Sa, Sb, Scd) = (0.269, 0.235, 0.177, 0.19, 0.098).  
Note that this is the mix in a magnitude limited catalog -- meaning 
that brighter galaxies (typically earlier-types) are over-represented.  


Figure~\ref{gmrMorph} compares the red sequence defined by our 
double-Gaussian fit to the color-magnitude relations defined by 
the different morphological types in the (significantly smaller) 
Fukugita et al. sample.  The top left panel shows that ellipticals 
do indeed lie along the same red sequence defined by the double-Gaussian 
fits; in particular, the steeper slopes at low and high luminosities, 
returned by our double-Gaussian fits to the full sample, are also 
evident in the smaller Fukugita et al. sample (see Huang \& Gu 2009 
for a more detailed analysis of the ``blue'' ellipticals with 
$g - r \leq 0.6$  -- they show either a star forming, AGN or 
post-starburst spectrum).  Thus, the curvature is not due to the 
fact that the mix of morphological types depends on luminosity.  

While S0s tend to define the same red sequence, a larger fraction 
are blue (top right).  The central panels show that types Sa and Sb 
can be both very red and very blue, and even types Sc and Sd can have 
rather red colors.  These red late-type galaxies are edge on disks; 
whereas any straight color cut will misleadingly group such objects 
together with early-types, the double-Gaussian decomposition correctly 
assigns these reddest objects at intermediate and low luminosities to 
the blue sequence.  


\section{Systematic effects}\label{systematics}


\begin{figure}
 \centering
 \includegraphics[width=0.99\hsize]{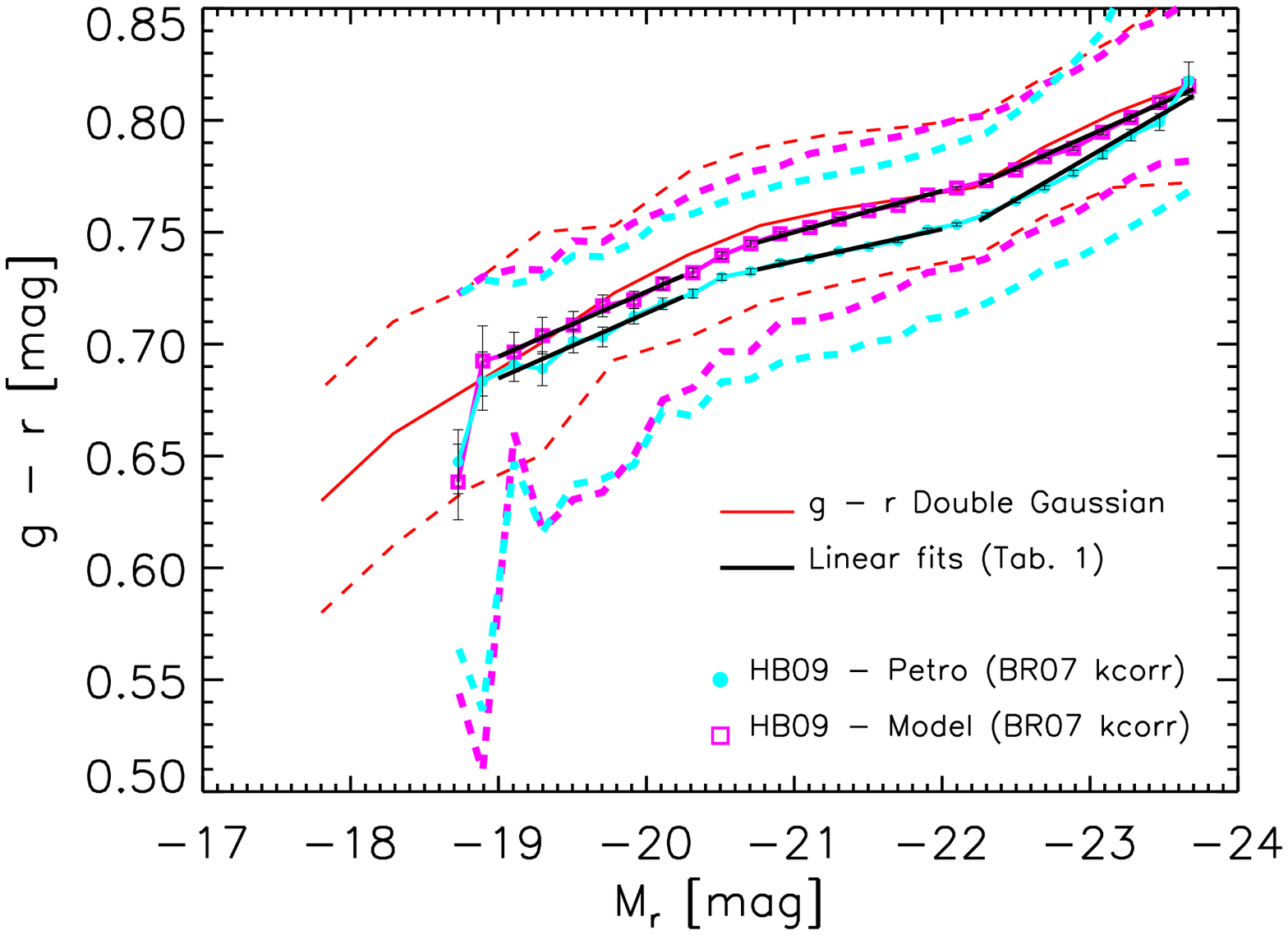}
 \caption{Dependence of color-magnitude relation on the scale on which 
          colors were defined. {\tt Model} colors, which have higher 
          signal-to-noise ratio, probe smaller scales, so are redder 
          than {\tt Petrosian} colors.  The color offset is largest at 
          $M_r\sim -22$, the luminosity scale at which color gradients 
          are maximal (Roche et al. 2010).}
 \label{gmrMPM}
\end{figure}

\begin{figure}
 \centering
 \includegraphics[width=0.95\hsize]{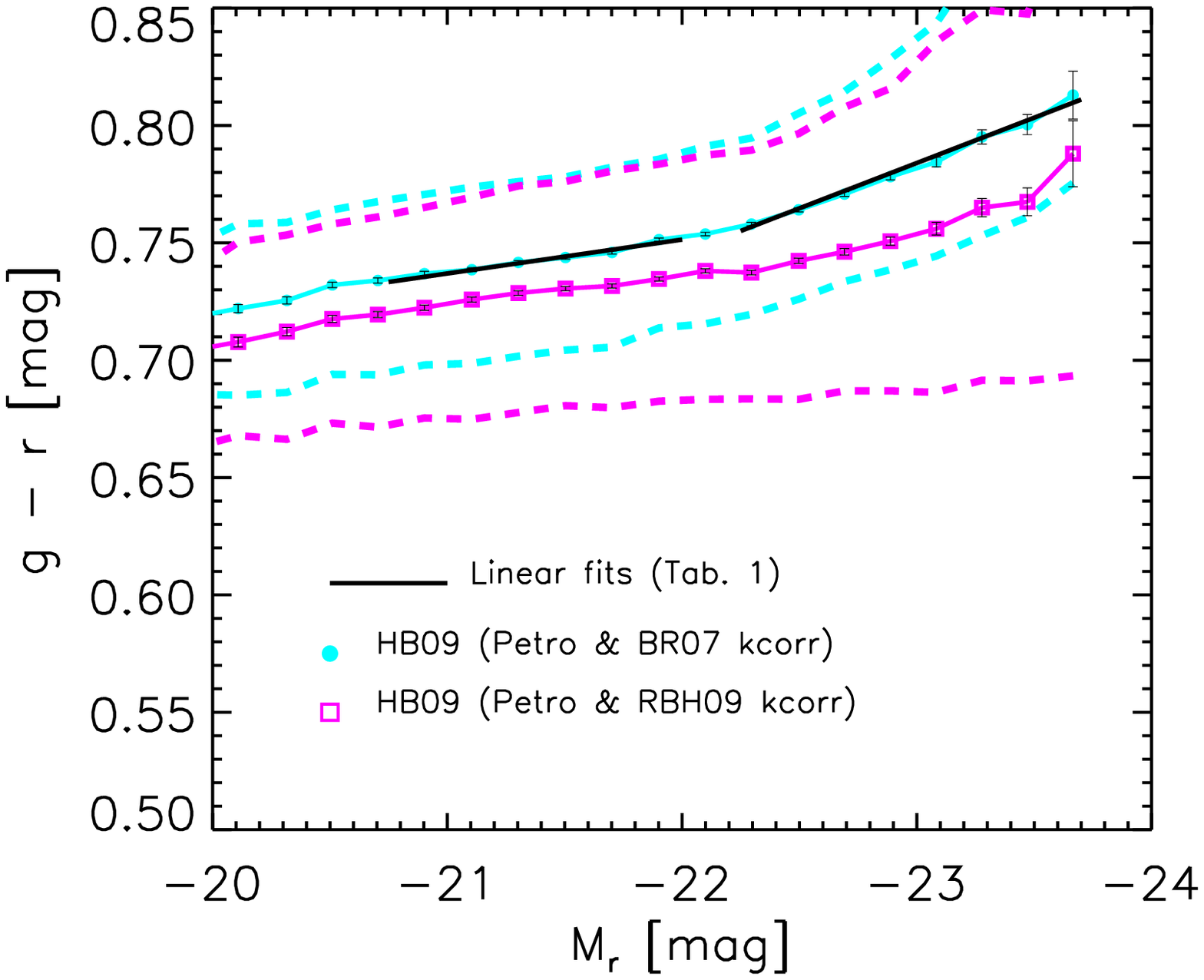}
 \caption{Dependence of color-magnitude relation on the choice of 
          $k$-correction: Blanton \& Roweis (2007; BR07) and 
          Roche et al. (2009; RBH09). Spectral-based $k$-corrections 
          (i.e. RBH09) appear to result in bluer colors and less curvature; 
          some of this is simply a consequence of the fact that the 
          spectra are taken using fibers of a fixed aperture.  }
 \label{gmrMPK}
\end{figure}

\begin{figure*}
 \centering
 \includegraphics[width=0.9\hsize]{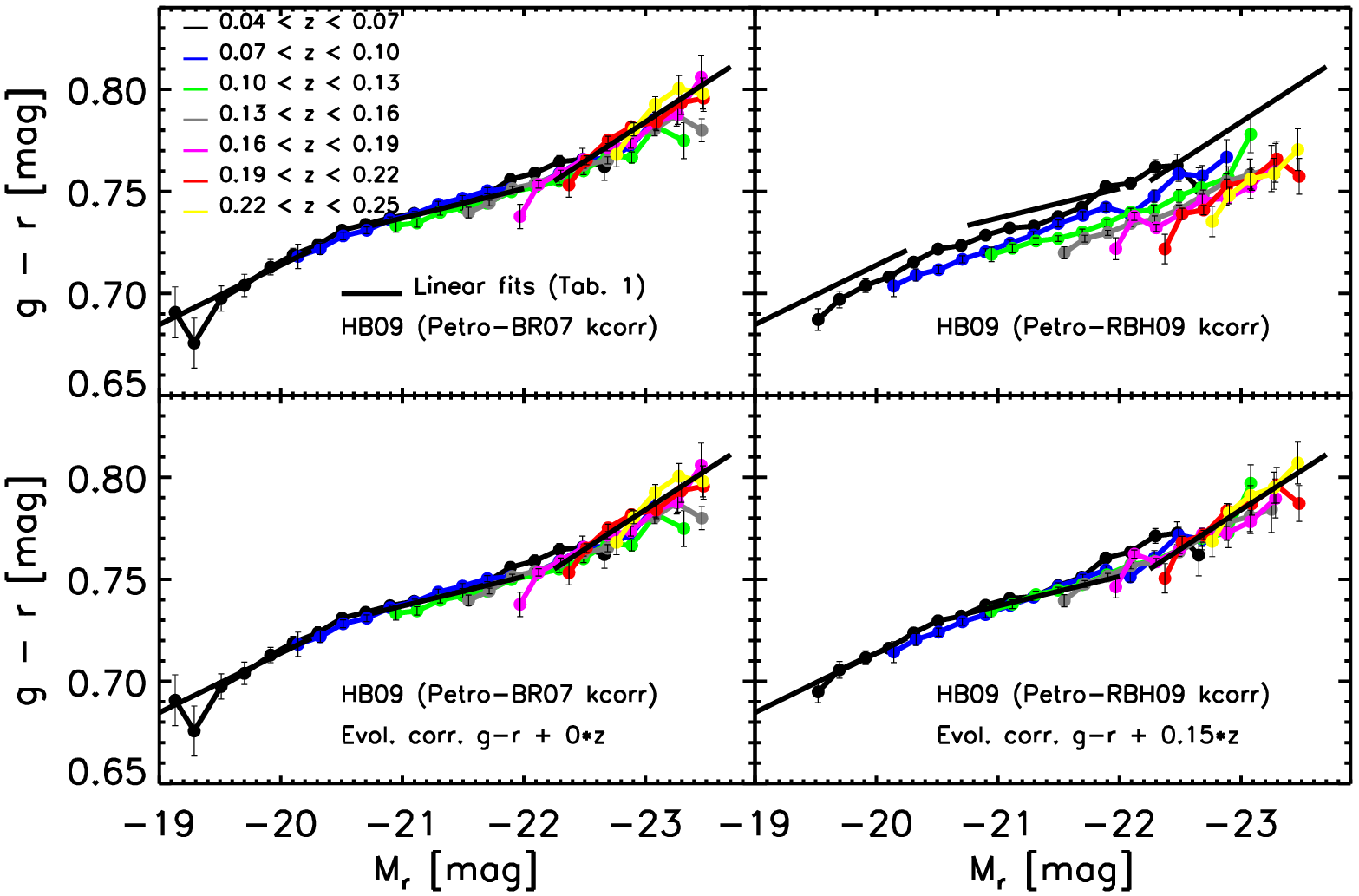}
 \caption{Dependence of color-magnitude relation on the choice of 
          $k$- and evolution corrections: 
          Blanton \& Roweis (2007; BR07 -- left panels) and 
          Roche et al. (2009; RBH09 -- right panels). 
          Different lines (colors) show
          the color-magnitude relation in different redshift bins as
          labeled.  No evolution correction has been applied in the 
          top panels.  Once luminosity evolution has been accounted for 
          (bottom panels) the curvature does not depend on the 
          $k$-correction (note that the evolution correction does 
          depend on the $k$-correction).  }
 \label{gmrEvolM}
\end{figure*}

\subsection{Effects of color gradients}\label{colgrad}
Because the curvature in the red sequence we reported in the main 
text is small, we have checked if it is robust to changes in how we 
estimate the colors and luminosities.  

The main text showed that the color-magnitude relation shows three 
distinct regimes (Table~\ref{gmrMtable} reports fits), whether one 
uses {\tt model} or {\tt Petrosian} quantities (compare 
Figures~\ref{gmrModel} and~\ref{gmrEtypes}), despite the fact that 
{\tt Petrosian} colors are slightly bluer than {\tt model} colors.  
The blueward shift occurs because the {\tt model} color probes the 
half-light radius, whereas the {\tt Petrosian} color is based on a 
larger physical scale, and early-type galaxies have negative color 
gradients (i.e., the are redder in the core; e.g. Wu et al. 2005).  
Although color gradients decrease with $\sigma$, they are a 
complicated function of luminosity: 
Gradients are largest for objects with $M_r = -22$, and are smaller 
for brighter or fainter objects (Roche et al. 2010).  
Figure~\ref{gmrMPM} shows a direct comparison:  the difference between 
the {\tt Petrosian} and {\tt model} color-magnitude relations in the 
Hyde-Bernardi sample is largest at $M_r\sim -22$ (note that the 
$M_r$ is the {\tt cmodel} quantity).  


For our purposes here, the main point is that three distinct regimes 
are seen whatever our choice of color, although it is interesting that 
they are slightly more obvious using colors which sample more of the 
total light of the galaxy:  at intermediate luminosities, the slope 
of the color-magnitude relation is flatter by a factor of two for 
{\tt Petrosian} rather than {\tt model} colors.  (The scatter around 
the mean relations is larger for Petrosian quantities, in part because 
of measurement errors -- recall from Section~\ref{data} that the model 
magnitudes are better measured.)

\subsection{Dependence on $k$- and evolution corrections}\label{k+e}
We have also tested for systematic effects which arise from $k$- 
and evolution corrections.  Our default has been to use values from 
Blanton \& Roweis (2007), which are based on fitting templates to the 
observed colors.  However, Roche et al. (2009) have recently described 
the results of estimating $k$-corrections from the spectra themselves.  
If we do not account for evolution, then the colors from the 
spectral-based $k$-corrections are slightly bluer at the bright end 
(Figure~\ref{gmrMPK}), resulting in weaker curvature. 
However, we have yet to account for luminosity evolution. 
The top panels in Figure~\ref{gmrEvolM} show the color-magnitude relation
in different redshift bins, before correcting for evolution, for the 
Blanton \& Roweis (left) and Roche et al. (right) $k$-corrections. 
(The plot uses {\tt Petrosian} colors, but the discussion is valid for 
the {\tt model} colors as well.) 
It is clear that we measure different evolution in the two cases: 
the evolution in $g-r$ is negligible when using the Blanton \& Roweis 
$k$-corrections (bottom left), while $g-r$ should be reddened by $0.15z$ 
for Roche et al. (bottom right). Once the color has been corrected for 
evolution in this way, the curvatures at the faint and bright ends are 
similar.

There is an additional subtle effect which arises from the fact that 
the spectra come from fibers having a fixed angular diameter of 3~arcsecs.  
Color gradients mean that the restframe light in the fiber from a higher 
redshift object will be slightly bluer, and this affects the 
spectral-based $k$-correction of Roche et al. (2009).  
In a magnitude limited sample, the more luminous objects are seen to 
higher redshifts, so this aperture effect can make the $k$-corrections 
masquerade as or erase curvature in the color-magnitude relation.  
The top right panel of Figure~\ref{gmrEvolM} also shows that if we 
restrict the sample to narrow redshift ranges, thus reducing both the 
evolution and simplifying aperture effects, the curvature in the 
color-magnitude diagram is still evident, at least in those bins where 
we have a sufficiently large range of luminosities.  (Of course, the 
significance of the curvature is smaller, because of the smaller sample 
sizes.)

This is important because  Hao et al. (2009) report that the slope of 
the color magnitude relation is steeper at $z\sim 0.3$ than at 
$z\sim 0.1$, and they interpret this as evolution in the slope of the 
relation.  We see this too -- the highest redshift samples (which span 
$M_r<-22.5$) appear to define steeper relations than those at $z<0.1$.  
However, because ours is a magnitude limited sample, these highest 
redshifts do not probe faint objects.  
Our lowest and intermediate redshift samples, which include a wider range 
of luminosities, show a slight upturn from intermediate to high luminosities, 
even at fixed redshift.  
Therefore, rather than concluding that the slope is 
evolving, we conclude that the slope depends on luminosity.  

\section{Sizes and velocity dispersions in zero-energy mergers}\label{0energy}
We assume that the final object is in virial equilibrium, 
that it formed from the merger of two smaller virialized objects 
in which mass was conserved, 
and that the total energy of the orbits which led to the merger 
was zero (sometimes called parabolic orbits).  

The virial condition means that $-W = 2K$ for all of the objects, 
meaning that the total energy for each object is 
$K+W = K - 2K = -K = W/2$.  If the mergers are of equal mass objects, 
each of mass $m$, then the total energy of the system before the 
merger is $-mv^2/2 - mv^2/2 = -2m\, (v^2/2)$ (because there is no 
contribution from the orbital energy).  However, the final object 
will have mass $2m$.  This with energy conservation and the 
constraint that the final object is virialized means that $v$ must 
be unchanged.   

If the mergers are not equal mass, then 
\begin{eqnarray}
 -\frac{mv^2}{2} - \frac{MV^2}{2} &=& - \frac{MV^2}{2} (f (v/V)^2 + 1) \\
  &=& -\frac{G (1+f)^2 M^2}{2R (R_f/R)}
   =  -(1+f)\,\frac{MV_f^2}{2},   \nonumber
\end{eqnarray}
where the larger object has mass $M$, size $R$ and velocity dispersion 
$V$, the smaller one has mass $m=fM$ and velocity dispersion $v\le V$, 
and $R_f$ and $V_f$ are the size and velocity dispersion of the 
final object which has mass $M(1+f)$.  
Eliminating a factor of $M(1+f)$ from the second and third expressions, 
and then using the fact that $-GM/2R = V^2/2$, gives an expression for 
$R_f/R$ in terms of $f$ and $v/V$:  
\begin{equation}
 \frac{R_f}{R} = \frac{(1+f)^2}{1 + f (v/V)^2}.
\end{equation}
In addition, equating the second and final expressions yields 
\begin{equation} 
 \frac{V_f^2}{V^2} = \frac{1 + f (v/V)^2}{1+f}.
\end{equation}
The density of the final object is proportional to 
\begin{equation}
 \frac{V_f^2}{R_f^2} = 
 \frac{V^2}{R^2} \frac{1 + f (v/V)^2}{1+f} \frac{[1 + f (v/V)^2]^2}{(1+f)^4}.
\end{equation}
Since $v/V\le 1$, the size will increase, and the velocity dispersion 
and density will both decrease.  The limiting case is when $f=1$: 
then $R_f = 2R$, $V_f=V$ and the density is smaller by a factor of 4.  
This is the basis for the claim that major mergers double the size without 
changing the velocity dispersion.  (Doubling the size decreases the 
density by a factor of 4 rather than $2^3=8$, because the mass has 
increased by a factor of 2.)
When $f\ll 1$ then $R_f/R \to 1 + 2f$ whereas $V_f/V \to 1 - f/2$:
for minor mergers, the fractional change to the size is larger than 
that to the velocity dispersion.  
The fractional change in density due to a minor merger is even larger:  
it scales as $-5f$.

\label{lastpage}

\end{document}